\documentclass[11pt,a4paper] {article}
\usepackage{amssymb,amsmath,amsfonts}
\usepackage[dvipsnames]{xcolor}
\usepackage{graphicx}
\usepackage{longtable}
\usepackage{verbatim}
\usepackage{color}
\usepackage[mathscr]{euscript}
\usepackage{framed}
\usepackage{mdframed}
\usepackage{amsmath}
\usepackage{amsfonts}
\usepackage{mathrsfs}
\usepackage{amssymb}
\usepackage{mathrsfs}  
\usepackage{subcaption}
\usepackage{caption}

\usepackage[left=26mm,top=20mm,right=26mm,bottom=20mm]{geometry}

\usepackage{diagbox}

\usepackage{color}
\usepackage{cite}
\usepackage{hyperref}
\hypersetup{colorlinks, citecolor=blu{\cal S}uro, linkcolor=black, urlcolor=bluscuro}
\definecolor{rossos}{cmyk}{0,1,1,0.55}
\definecolor{blu{\cal S}uro}{rgb}{0.15, 0.2, .85}
\definecolor{bluchiaro}{cmyk}{1,.3,0.,0.1}

\graphicspath{{./Figures/}}

\newcommand{\be}{\begin{equation}}
\newcommand{\ee}{\end{equation}}
\newcommand{\bea}{\begin{eqnarray}}
\newcommand{\eea}{\end{eqnarray}}
\newcommand{\beq}{\begin{equation}}
\newcommand{\eeq}{\end{equation}}

\def\beqa{\begin{eqnarray}}

\def\eeqa{\end{eqnarray}}

\def\lsim{\mathrel{\rlap{\lower4pt\hbox{\hskip0.5pt$\sim$}}
    \raise1pt\hbox{$<$}}}         
\def\gsim{\mathrel{\rlap{\lower4pt\hbox{\hskip0.5pt$\sim$}}
    \raise1pt\hbox{$>$}}}         

\usepackage[normalem]{ulem}

\makeatletter
\renewcommand{\citation}[1]{%
  \g@addto@macro{\citation@list}{,#1}%
}
\newcommand*{\citation@list}{} 
\newcommand{\sortbibitem}[2]{%
  \global\@namedef{bibitem@#1}{%
    \bibitem{#1} #2
  }%
}
\newcommand{\sort@bibitems}{%
  \@for\next:=\citation@list\do{%
    \@nameuse{bibitem@\next}%
    \global\@namedef{bibitem@\next}{}%
  }%
}
\expandafter\def\expandafter\endthebibliography\expandafter{%
  \expandafter\sort@bibitems\endthebibliography
}
\makeatother

\begin{document}

\vspace{0.1in}

\begin{center}
{\Large\bf\color{black}
Features in the Inflaton Potential
and
the Spectrum of Cosmological Perturbations
}\\
\bigskip\color{black}
\vspace{.5cm}
{ {\large  \rm  Ioannis Dalianis} 
\vspace{0.3cm}
} \\[5mm]
{\it {Department of Physics, University of Cyprus, \\
Nicosia 1678, Cyprus \\ 
and \\
National and Kapodistrian University of Athens, Department of Physics, \\
 Section of Nuclear {\rm \&} Particle Physics,  GR--157 84 Athens, Greece
}}\\[2mm]

\end{center}

\vskip.2in

\noindent
\rule{16.6cm}{0.4pt}

\vspace{.3cm}
\noindent
{\bf \large {Abstract}}
\vskip.15in 

Cosmological perturbations, originating in the quantum fluctuations of the
fields that drive inflation, are observed to be nearly scale invariant
at the largest scales. At smaller scales, however, perturbations are not
severely constrained and might be of particular importance if their amplitude is
large. They can trigger the creation of primordial black holes (PBHs) or stochastic
gravitational waves (GWs). Small-scale perturbations are generated during the
later stages of inflation, when possible strong features in the inflaton
potential can break scale invariance and leave characteristic imprints on the spectrum. 
We focus on and review three types of features: inflection points and steep steps
in the potential, as well as sharp turns in the inflationary trajectory in field
space. We show that such features induce a strong enhancement of the curvature
spectrum within a certain wavenumber range. In particular cases, they also generate
characteristic oscillatory patterns that are transferred in the spectrum of secondary GWs, 
which are potentially observable by operating or designed experiments. We demonstrate these effects
through the calculation of the primordial power spectrum and the PBH
abundance in the context of $\alpha$-attractors and supergravity (SUGRA) models of
inflation.

\noindent

\bigskip

\noindent
\rule{16.6cm}{0.4pt}

\vskip.4in


\section{Introduction}

Inflationary  models that feature strong deviations from the slow-roll (SR) conditions can  produce a scale dependent power spectrum of fluctuations. A particularly interesting case is that of an amplified spectrum at wavelengths smaller than that of the CMB scales, where 
the bounds on the primordial spectrum of curvature perturbations $\Delta^2_{\cal R}(k)$ are rather weak, see e.g. \cite{Carr:2020gox, Carr:2017edp, Dalianis:2018ymb, Gow:2020bzo, Yang:2020egn, Ando:2022tpj}.
The motivation to examine such a spectrum is twofold. Firstly, it  is an observationally viable possibility with essential cosmological implications, such as
the existence 
of secondary gravitational waves (GWs) \cite{Matarrese:1992rp, Matarrese:1997ay, Noh:2004bc, Carbone:2004iv, Ananda:2006af, Baumann:2007zm}
 and, possibly, 
of primordial black holes (PBH) \cite{Carr:1974nx, Meszaros:1974tb, Novikov}.
  Secondly, it points towards new and challenging inflationary model building directions that go beyond the standard paradigm. 
Notable inflationary mechanisms that can realize a sizable amplification display the following features:
\begin{itemize}

\item  step-like  transitions in the inflaton potential energy,

\item approximate inflection points, 

\item sharp turns in the multi-field space of the inflaton potential.

\end{itemize}
These mechanisms are associated with a strong  acceleration and deceleration  of the inflaton field. 
In the context of canonical inflation 
such  changes in the velocity
are attributed to strong features in the inflationary potential, especially designed and located, see e.g. ref. \cite{Ozsoy:2023ryl} for a recent discussion.
 
A {\it step-like} feature 
causes a fast decrease of the potential energy and  produces spectral distortions \cite{Starobinsky:1992ts, Adams:2001vc, Kefala:2020xsx, Dalianis:2021iig,  Inomata:2021tpx, Pi:2022zxs}:  One or more  transition points in the inflaton  potential  energy, at which it jumps from one constant value to another, can result in a sizable  enhancement of the spectrum within a certain range of  scales.  
On top of that, characteristic oscillatory spectral patterns are produced.
Spectral enhancement due to the {\it inflection point} mechanism requires that the inflaton traverses a region of its potential $V$  with  $V'\sim V'' \sim 0$ \cite{Ivanov:1994pa}. This is a local flattening of the potential that might be deformed so that  a  shallow minimum and maximum appears. 
In that region the inflaton gains and loses kinetic energy, which results in an enhancement of the primordial spectrum.  Several models implementing this idea  
have been put forward in the last couple of years, see e.g. \cite{Garcia-Bellido:2017mdw, Kannike:2017bxn,  Germani:2017bcs,  Motohashi:2017kbs,  Ballesteros:2017fsr, Hertzberg:2017dkh, Cicoli:2018asa, Ozsoy:2018flq, Dalianis:2018frf} for some first works. 
In multi-field inflationary dynamics a different source for intense superhorizon evolution of the curvature perturbations can exist. 
The mechanism is associated with a {\it sharp turn} where a   slow-roll violation along a direction orthogonal to the inflationary trajectory takes place. 
This transient 
phase amplifies the amplitude of the  otherwise catatonic isocurvature modes.
The amplification of isocurvature perturbations  is converted through suitable couplings into an amplification of curvature perturbations \cite{Palma:2020ejf, Fumagalli:2020nvq, Fumagalli:2020adf, Braglia:2020eai, Braglia:2020taf, Fumagalli:2021mpc, Boutivas:2022qtl}.

\section{Beyond slow-roll single-field inflation}

In the framework of the inflationary universe, the cosmological perturbations naturally
emerge from quantum zero-point fluctuations \cite{Mukhanov:1981xt}.
Fluctuations of the inflaton scalar field $\phi$ on top of a homogeneous background  induce scalar perturbations in the metric, which  can be described through the
comoving gauge curvature perturbation ${\cal R}$.
The variance in the fluctuations is quantified by the power spectrum $\Delta^2_{\cal R}(k)$.
Expanding the inflaton-gravity action for canonical single field inflation
\begin{equation}
S= \int d^4x \sqrt{-g}\left[\frac12 R + \frac12 (\partial \phi)^2 - V(\phi)\right]\,
\end{equation}
up to second order in ${\cal R}$, one obtains
\begin{equation}
S_{(2)}= \frac12 \int {\rm d}^4x \, a^3 \, \frac{\dot{\phi}^2}{H^2} \left[\dot{{\cal R}}^2 -\frac{(\partial_i {\cal R})^2}{a^2}\right]\,,
\end{equation}
 where  overdots denote derivatives with respect to the coordinate time $t$.  We have chosen units such that the reduced Planck mass is  $M_{\rm Pl}=1$.
After the variable redefinition $v=z{\cal R}$ where $z^2=a^2\dot{\phi}^2/H^2=2a^2\epsilon_1$  and switching to conformal time $\tau$, defined by $d\tau=dt/a$, the action is recast into
 canonical  form ($\epsilon_1$ is defined in eqs. (\ref{eek})).
The evolution of the Fourier modes $v_k$  are described by the Mukhanov-Sasaki equation
\begin{equation} \label{eqMS}
v''_k +\left(k^2-\frac{z''}{z}\right) v_k =0 ,
\end{equation}
where the potential term $z''/z$ is exactly expressed in terms of the Hubble flow functions  
\begin{equation} \label{eek}
\epsilon_1 \equiv -\frac{\dot{H}}{H^2} =\frac{\dot{\phi}^2}{2H^2}, \quad
 \epsilon_2 \equiv \frac{\dot{\epsilon}_1}{H\epsilon_1}=2\epsilon_1+2\frac{\ddot{\phi}}{H\dot{\phi}}, \quad 
 \epsilon_3 \equiv \frac{{\dot{\epsilon}}_2}{H\epsilon_2}, 
\end{equation}
as 
\begin{equation} \label{zz}
\frac{z''}{z} =(aH)^2 \left[2-\epsilon_1 +\frac32 \epsilon_2 - \frac12\epsilon_1 \epsilon_2 +\frac14 \epsilon^2_2+\frac{1}{2} \epsilon_2 \epsilon_3 \right]. 
\end{equation}
Here the prime denotes derivatives with respect to the conformal time $\tau$.
An alternative to $\epsilon_2$ is the the second slow-roll parameter $\eta=\epsilon_1-\epsilon_2/2$ that we will use  later (section \ref{sec-turns}).
The super-Hubble evolution of the curvature perturbation, i.e. $k^2 \ll z''/z$,
 is a linear combination of $z$ and $z\int d\tau/z^2$.  Given that ${\cal R}_k=v_k/z$ one finds 
\begin{equation} \label{Rsr}
{\cal R}_k =c_1(k) +c_2(k) \int \frac{dt}{a^3 \epsilon_1} \,,
\end{equation}
where $c_1$ and $c_2$ are  integration constants. 
The curvature power spectrum is defined as
\begin{equation}
\Delta^2_{\cal R}=\frac{k^3}{2\pi^2} \,|{\cal R}_k|^2\,
\end{equation}
and is estimated at a time  after the mode exits the horizon and its value freezes.

\begin{figure}[t!]
\centering
\includegraphics[width=0.47\textwidth]{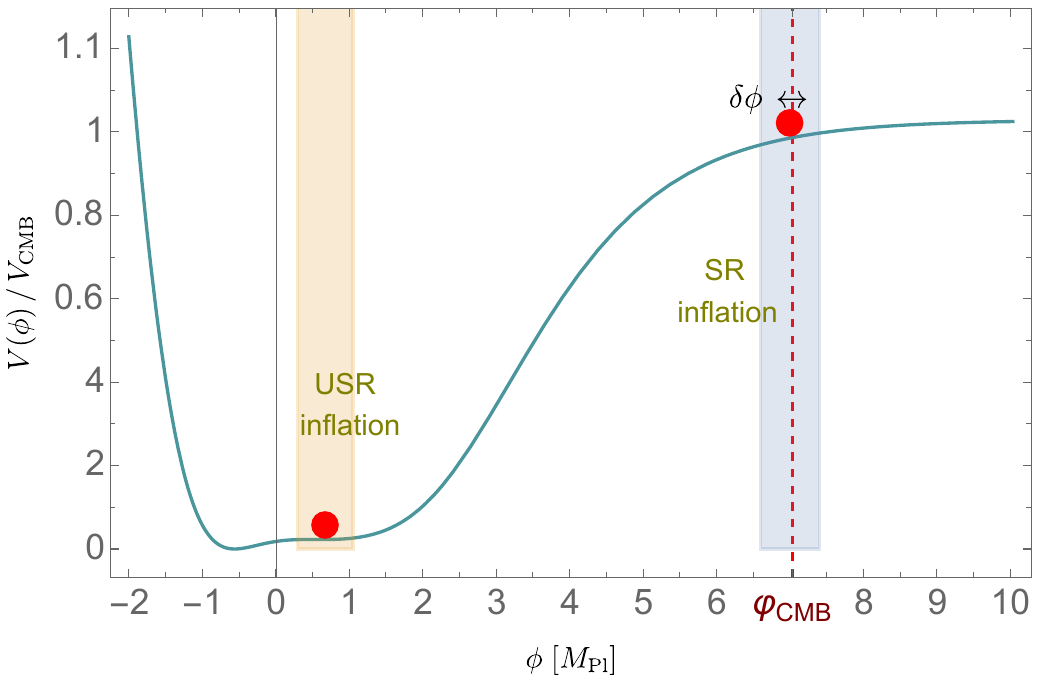}
\hspace{0.3cm} 
 \includegraphics[width=0.47\textwidth]{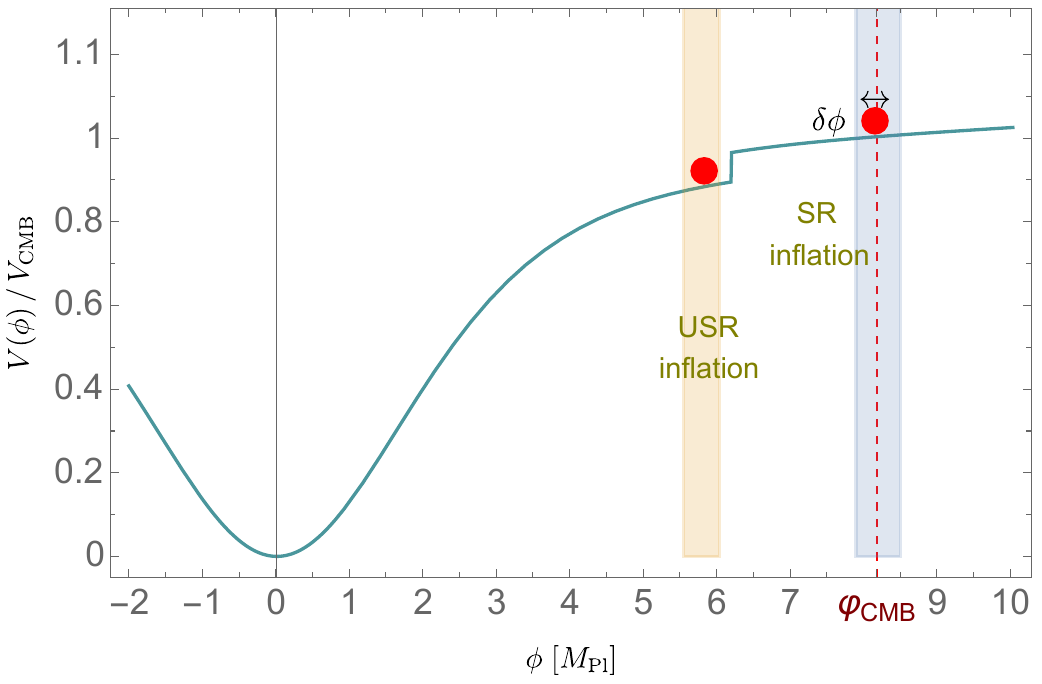} 
\includegraphics[width=0.65\textwidth]{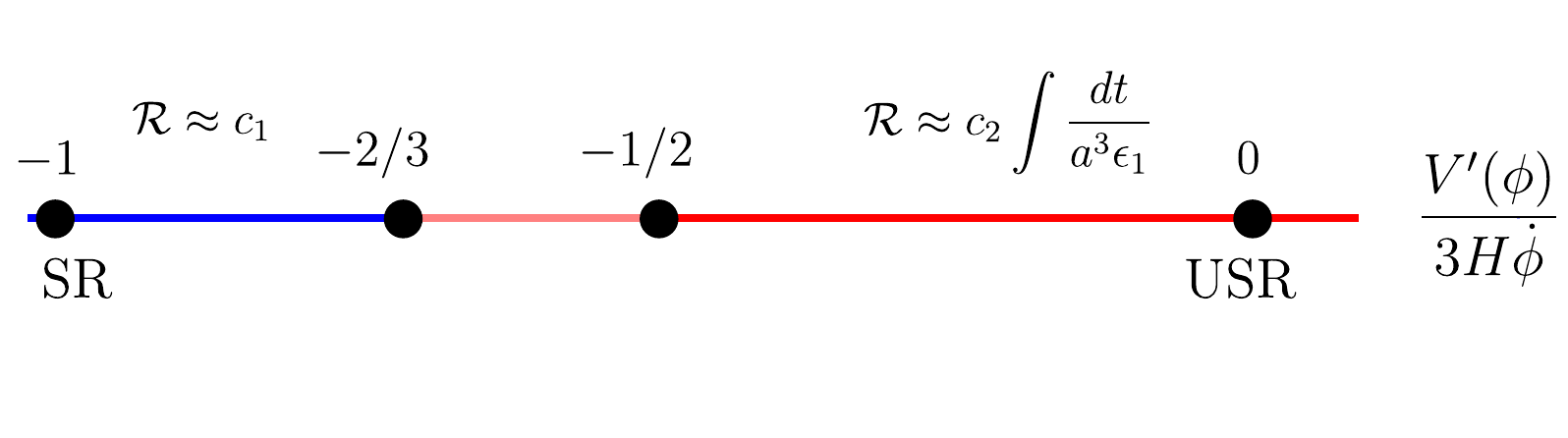}
\caption{
{\it Upper panels:} Two examples of inflationary potentials in the framework of $\alpha$-attractors that realize both a SR and an USR phase: one with an infection point (left panel) and one with a step-like tranistion (right panel). 
{\it Lower:} 
Intervals  of the ratio $V'/(3H\dot{\phi})$ within which  a SR or USR phase takes place. The solution for ${\cal R}_k$ that characterizes each interval is also shown.
}
\label{graphs}
\end{figure}

 In the conventional single-field inflationary scenarios based on the slow-roll analysis, the second term in  eq. (\ref{Rsr})
  correspond to a decaying contribution, so that  ${\cal R}_k$ soon becomes a constant after  horizon crossing. 
For a nearly constant ratio $V'(\phi)/(3H\dot{\phi})$ 
the first Hubble flow parameter  can be found analytically \cite{Motohashi:2014ppa} and scales as $\epsilon_1\propto a^{2 \delta}$, 
 where
\begin{equation} \label{delta}
\delta\equiv\frac{\ddot{\phi}}{H \dot{\phi}}=-3\left( 1+\frac{V'(\phi)}{3H \dot{\phi}}\right)\,.
\end{equation}
For $\epsilon_1 \ll 1$ we have $2\delta\approx \epsilon_2$.  
During slow-roll inflation we have $\epsilon_2 \ll1$ and 
the second term in  eq. (\ref{Rsr}) is a decaying mode, since  $\epsilon_1\propto a^{ \epsilon_2}$ and ${\cal R}_k$ is dominated by the constant term. 
The power spectrum can be described quantitatively well by the expression  $\Delta^2_{\cal R} \approx H^2/(8\pi^2 \epsilon_1)$ during the slow-roll regime. The dominant solution is a constant mode and the  curvature perturbation is conserved outside the Hubble radius. 

The other limit is the ultra-slow roll (USR) inflation \cite{Kinney:2005vj}, which corresponds to a regime in which 
the effect of the potential is negligible in the equation of motion  of the inflaton.
In this regime,
 the Klein-Gordon equation takes  the form $\ddot{\phi}\approx -3H \dot{\phi} $, and 
 from eq. (\ref{delta}) we see that $\epsilon_2 \approx -6$. Therefore, the first Hubble-flow parameter decreases rapidly, i.e.
 $\epsilon_1=\dot{\phi}^2/(2 M^2_{\rm Pl} H^2)\propto a^{-6}$,  and  the solution (\ref{Rsr}) is dominated by the second mode which grows as $(a^3 \epsilon_1)^{-1} \propto  a^3$.    
 These two limits\footnote{Values $-6 \leq  \epsilon_2 \leq -3$ correspond to a phase called non-single-clock inflation \cite{Byrnes:2018txb}.} are sketched in the graph of fig. \ref{graphs}.

During USR the scalar field moves on a flat potential. It may initially fast-roll and then decelerate, with  the slow-roll approximation breaking down, but it eventually stops because of  the Hubble friction at a point that  depends on its initial condition.
USR inflation has a graceful exit problem and it is also incompatible with the measured tilt of the CMB power spectrum. 
 However, an USR phase is a viable possibility if it begins after the CMB modes exit the horizon, as long as it is guaranteed  that the CMB modes do not grow after horizon exit \cite{Kristiano:2022maq, Riotto:2023hoz}. The USR phase has to be followed by a slow-roll or a constant-roll phase \cite{Motohashi:2014ppa}. 
The motivation to introduce an USR phase is so that it greatly enhances the scalar power spectrum in single-field inflationary models.
The enhancement is indicated already by the
slow-roll result $\Delta^2_{\cal R}\propto 1/\epsilon_1$, which increases for decreasing $\epsilon_1$  values, 
but 
 this result is not valid during USR and does not  describe  accurately the amplification factor  or the characteristic wavenumbers.
The correct picture is given
by the full solution, eq.  (\ref{Rsr}), that indicates a more subtle  superhorizon evolution for the curvature perturbation \cite{Leach:2000yw,  Leach:2001zf, Saito:2008em}.

\subsection{Amplification}

An USR is  phase is driven by a nearly flat part of the potential,   where  $V'(\phi)$ is negligible compared to the acceleration $\ddot{\phi}$ and the friction term $3H\dot{\phi}$ in the equation of motion.
We distinguish the following cases:
\begin{itemize}

\item The first case is that of 
a monotonic potential  with  a positive derivative, i.e. $V'(\phi)> 0$,  where the inflaton  continuously rolls down towards smaller potential values.
The presence of an USR regime after a slow-roll  slope, with strong acceleration followed by deceleration,   implies a transition that is connected to the presence of a step in the inflaton potential. 
In refs. \cite{Kefala:2020xsx, Dalianis:2021iig} it was pointed out that the steepness of this step-like transition  is critical, with sharp steps enhancing  $\Delta^2_{\cal R}$ significantly.

\item The second case is that of a non-monotonic potential 
with  $V'(\phi)$ changing sign. 
The conventional example  is that of an approximate inflection point in the potential. Literally, an inflection point means that there is a field value at which $V'(\phi)=V''(\phi)=0$. However,  a stronger amplification of  $\Delta^2_{\cal R}$ is achieved if the region around the inflection point  is deformed into  a local minimum and maximum.  In this sense, an approximate inflection point  indicates that a shallow potential well separates two slow-roll phases and realizes a transient USR regime. 

\end{itemize}

\begin{figure}[t!]   
  \begin{subfigure}{.5\textwidth}
  \centering
  \includegraphics[width=1.0 \linewidth]{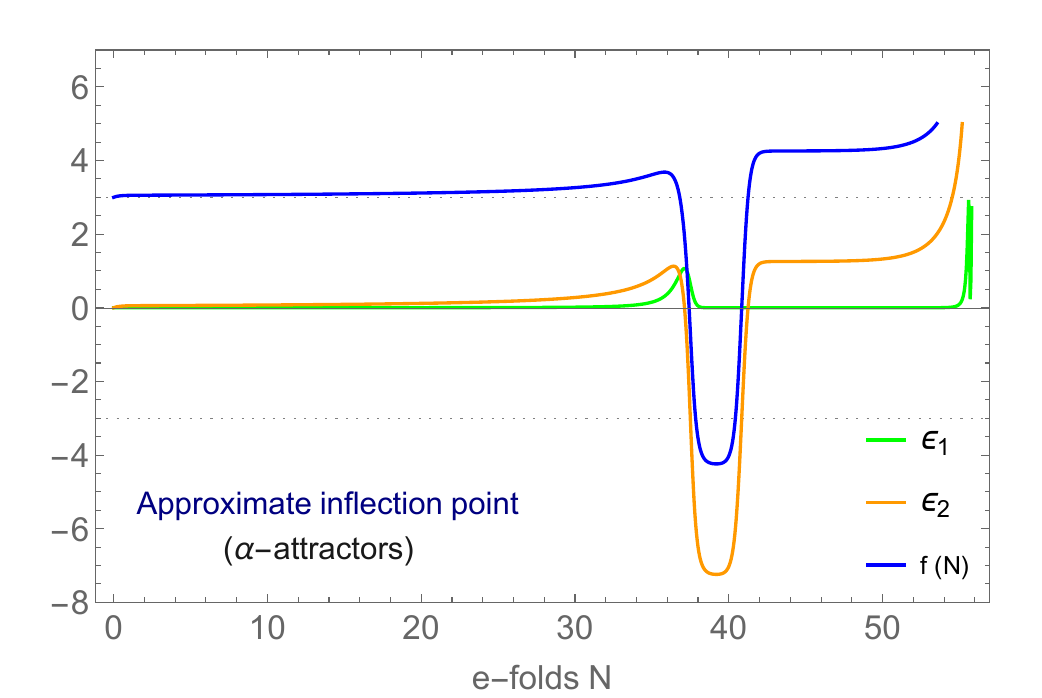}
\end{subfigure}  \quad\quad
  \begin{subfigure}{.5\textwidth}
  \centering
  \includegraphics[width=1\linewidth]{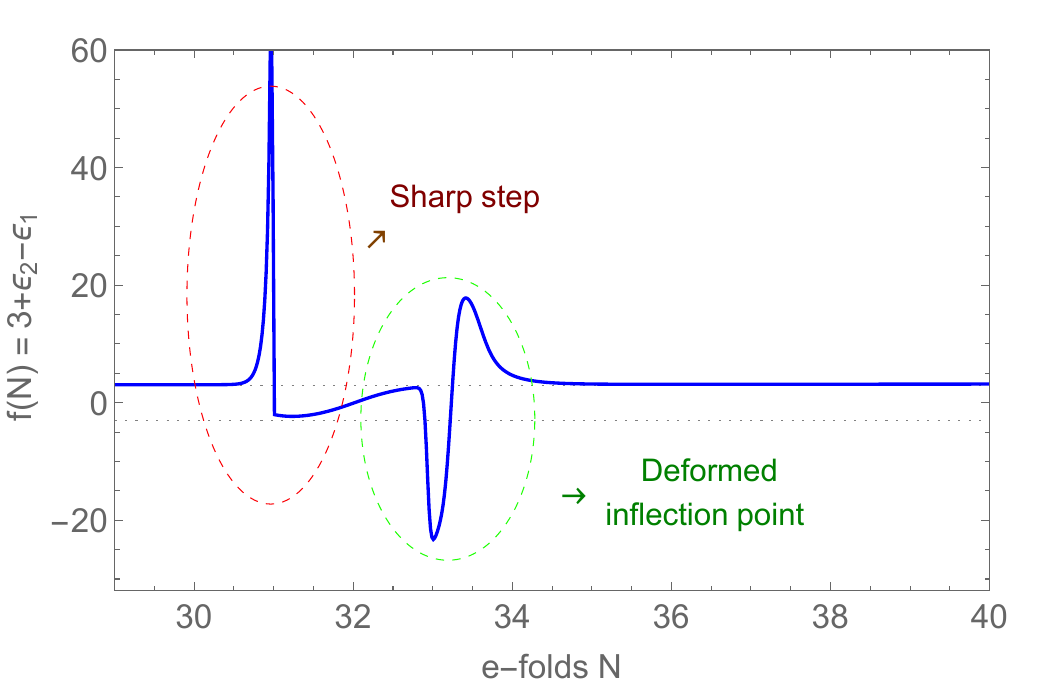}
\end{subfigure}
 \caption{\label{f2}~~  
{\it Left  panel}: The evolution of the first two Hubble flow parameters, $\epsilon_1$ (green), $\epsilon_2$ (orange), and the function $f(N)$ of eq. (\ref{AN}) for an inflationary model in the context of $\alpha$-attractors \cite{Dalianis:2018frf}.  
{\it Right Panel}: The evolution of the function $f(N)$ for an inflationary model with  a sharp step followed by an inflection point \cite{Dalianis:2021iig}. In both plots the dotted grid-lines indicate the critical values $\pm 3$ for the function $f(N)$.
}
\end{figure}

In fig. \ref{f2} a comparison  between the aforementioned mechanisms is demonstrated in terms of the $\epsilon_1$ and $\epsilon_2$ parameters. In the left panel of fig. \ref{f2} the inflection point is approximate, i.e. there is a smooth shallow minimum and maximum. In the right panel there is a combination of a step and an  inflection point. We display this non-standard inflection point because it is characterized by  a sharp minimum and maximum,  for which the potential contribution is important.  For a step-like feature there is a short phase of very strong acceleration which is followed by  a period of deceleration 
when the inflaton settles onto the second plateau.
For the inflection point the most prominent feature is the deceleration.

\begin{itemize}
\item A third case of $\Delta^2_{\cal R}$ amplification, which will be discussed in a following section, is that of perpendicular  acceleration in multi-field inflation. 
\end{itemize}

Also, we comment that relatively slow turns which appear in hybrid inflation models  \cite{Linde:1993cn} can  lead to a change of the inflationary
regime and amplification of small scale perturbations\cite{Garcia-Bellido:1996mdl}.
 Particular set ups of  related mechanisms as well as other model building directions  can be found in refs. \cite{DAmico:2020euu, Spanos:2021hpk,  Ballesteros:2022hjk, Braglia:2022phb, Choudhury:2023hvf}, just to mention a few.

\subsubsection{An analytic calculation}

We would like to have  a  quantitative  description of the increase of the amplitude of the curvature perturbation. 
We start from the equation for the curvature perturbation in Fourier space,
\be
{\cal R}_k''+2\frac{z'}{z}{\cal R}_k'+k^2{\cal R}_k=0 \,.
\label{geqqf} 
\ee
In terms of efoldings $N$
 the equation for the curvature perturbation takes the form
\be
R_{k,NN}+f(N)\, R_{k,N}+\frac{k^2}{e^{2N} H^2} R_{k}=0\,,
\label{RN} 
\ee
where the parameter $z$ reads  $z= e^N\, \phi_{,N}$ 
and the subscript $N$ indicates a derivative.
The solutions for ${\cal R}_k$ depend on the quantity 
\be
f(N)=3+\frac{2\phi_{,NN}}{\phi_{,N}}-\frac{\phi_{,N}^2}{2 M_{\rm Pl}^2}=3+ \epsilon_2 -\epsilon_1\,.
\label{AN} \ee
In the slow-roll regime,
$f(N)$ acts as a generalized friction term. However, if $\epsilon_2$ becomes 
negative,  it acts in an opposite way and can lead to
a dramatic enhancement of the perturbations.

Within the slow-roll  approximation, we have $f(N)\approx 3$ and 
the solution of eq. (\ref{RN}) can be expressed in terms of the 
Bessel functions  $J_{\pm 3/2}$ as
\be
{\cal R}_k(N;C_{p},C_{m},3)=A \, e^{-\frac{3}{2}\, N} 
\left(C_{p}J_{3/2}\left( e^{- N}\frac{k}{H} \right) + C_{m}\, J_{-3/2}
\left( e^{-N}\frac{k}{H} \right)  \right),
\label{solBessel3} \ee
where we take $A$ to be real without loss of generality.
For 
$-k\tau \gg 1$ we assume the  Bunch-Davies vacuum as the initial condition for the evolution of the fluctuations. This selects  
 the values $C_p=1$, $C_m=i$. The curvature pertubation is
${\cal R}_k(N;1,i,3)\propto \left( e^{-i k \tau}/\sqrt{k} \right) \, (1-i/(k\tau))/ a(\tau)$, 
where the
conformal time is $H \tau=-e^{-N}=-1/a$.  
 Within slow-roll and for late times $\tau \to 0^-$,   the curvature perturbation approaches a constant value 
 as the 
mode ${\cal R}_k$ moves out of the horizon and freezes, with  the absolute value of $C_m$  determining the value of the power spectrum.

This simple picture is modified when the function $f(N)$ deviates from a constant value
equal to 3, or equivalently when $|\epsilon_2|$ becomes large. 
This can occur during an USR phase, 
 where the spectrum might be enhanced by several orders of magnitude and new analytical techniques are needed.
Generally, we are interested in  one or more transient {\it non-slow-roll}  phases that take place between an initial and a final slow-roll phase. 
During such a phase
the  velocity $\varphi,_{N}$ changes fast, by growing or decaying depending on the sign of $f(N)-3.$
In order to obtain an analytical solution and quantitative understanding, 
we model $f(N)$ through a sequence of square ``pulses", each with constant $f(N)=\bar{f}_i\not=3$. 
We also approximate the Hubble parameter $H$ as almost constant, which  is justified  because the change in the Hubble parameter for an
inflection point or a step is at most of order $10\%$, while the enhancement of the spectrum can be several orders of magnitude large. We note that a similar analysis and supplementary results can be found in refs. \cite{Byrnes:2018txb, Carrilho:2019oqg, Ng:2021hll}.

\begin{figure}[t!]
\centering
\includegraphics[width=0.4\textwidth]{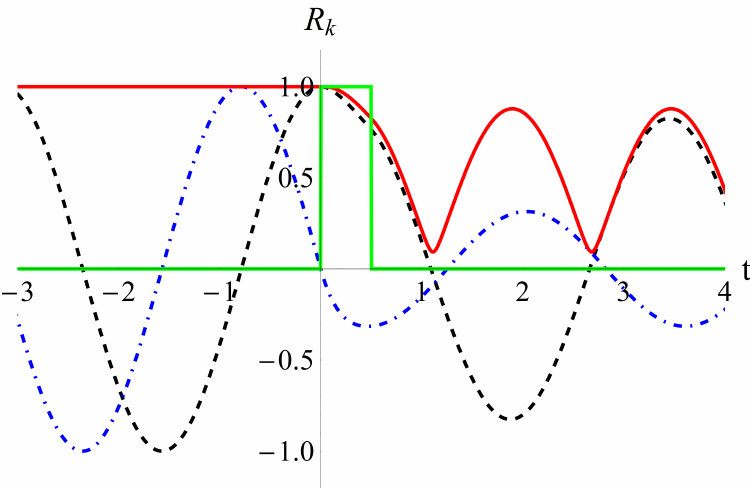}
\hspace{0.5cm}
 \includegraphics[width=0.4\textwidth]{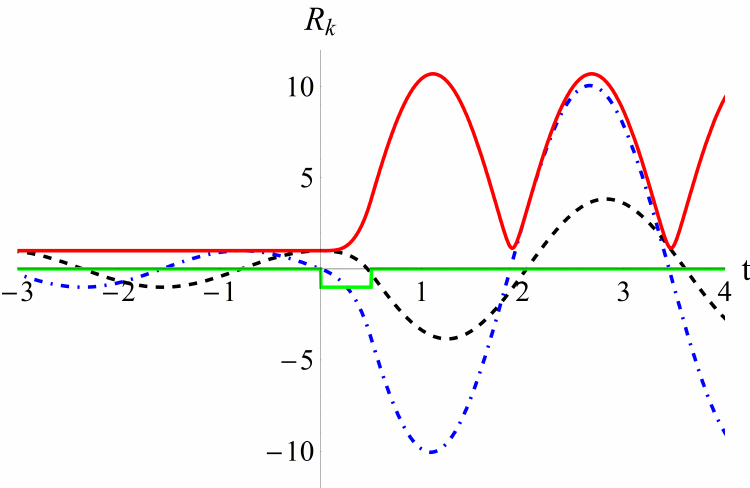} 
\caption{
The real part (dashed curve), imaginary part (dot-dashed curve) and 
amplitude (red solid curve) of ${\cal R}_k(t)$ for the toy-model equation ${\cal R}_{k,tt}+\bar{f}{\cal R}_{k,t}+ k^2 {\cal R}_{k}=0$ for a ``pulse" in 
the interval $0\leq t \leq 0.5$ and with $k=2$. 
We also display the ``pulse'', with a rescaled
maximum $\bar{f}/5$.
{\it Left panel}: $\bar{f}=5$. 
{\it Right panel}: $\bar{f}=-5$ \cite{Dalianis:2021iig}. 
}
\label{plotwave1}
\end{figure}
 
Our starting point is the solution eq. (\ref{solBessel3}). 
 We assume that a strong feature appears at some moment in time.  
 With the term feature we mean sharp deviations from the minimal and smooth inflationary evolution that have observable consequences, see e.g. refs.  \cite{Chen:2010xka, Chluba:2015bqa, Slosar:2019gvt}. 
   Here we focus on features   that can be approximated with a top-hat function that we call  a square ``pulse".
  The motivation behind  this  approximation is
the geometrical shape of the function $f(N)$ when a strong feature, such as an inflection point or a step, affects the inflationary  trajectory, as depicted in fig. \ref{f2}.
For $k\rightarrow 0$  the evolution of ${\cal R}_k$  is given by 
 eq. (\ref{solBessel3}), resulting in 
 the asymptotic 
value $|C_m|=1$ that corresponds to a scale-invariant spectrum, normalized to the CMB value.
We are interested in the relative increase of the asymptotic value of $|C_m|$ 
if at some value of $N$ 
a ``pulse" appears in $f(N)$.
 For a square ``pulse" of height $\bar{f} \neq 3$ 
 that begins at $N_1$ and ends at $N_2$, we have
\begin{equation}
 \epsilon_2(N)=(\bar{f}-3)\left[\Theta(N-N_1) - \Theta(N-N_2) \right]\,,
\end{equation}  
where $\Theta$ is a step function. The solution involves a linear combination of the Bessel functions $J_{\pm \bar{f}/2}$ and  $J_{\pm 3/2}$.
The corresponding values of the  constants $C'_{p}$, $ C'_{m}$ after the pulse is traversed  are given by the matrix equation \cite{Dalianis:2021iig}
 \be
\begin{pmatrix}  
		 C'_{p}\\
		  C'_{m}
\end{pmatrix}=M_{\rm pulse}(N_1, N_2,\bar{f}, 3,k)
\begin{pmatrix}	  
		1\\
		  i
\end{pmatrix}\,.
\label{coefficients}	  
\ee 
The matrix $M_\text{pulse}$ is a product of two square matrices,  
 with each one giving the new  coefficients $(C_{p}, C_{m})$ for the transitions $3 \rightarrow \bar{f}$ at $N=N_1$ and $ \bar{f} \rightarrow 3$ at $N=N_2$.
 The two transitions combined account for the effect of a single square ``pulse". 
 The matrix $M_\text{pulse}$ 
has  elements that depend on the Bessel functions $J_{\bar{f}/2}$, $J_{-\bar{f}/2}$, $ J_{1-\bar{f}/2}$, $ J_{-1+\bar{f}/2}$ and  $J_{3/2}$, $ J_{-3/2}$, $ J_{1/2}$, $ J_{-1/2}$, which
 can be found by requiring the continuity of the solution and its first derivative at $N = N_1$ and $N=N_2$.

A more general set up involves the evolution of the fluctuation ${\cal R}_k$ 
 through a sequence of strong features  parameterized by a series of ``pulses" in $f(N)$. 
  A product of several $M_{\rm pulse}$ matrices can reproduce the final values of the coefficients $(C'_{p}, C'_{m})$ of the Bessel functions. In the case of a final slow-roll regime after the ``pulses", the Bessel functions for large $N$ are  $J_{\pm 3/2}$. 
  Clearly, it is possible to reconstruct any smooth function $f(N)$ in terms of short intervals of $N$ during which the function takes constant values.  
For a single pulse we obtain that for $\bar{f} >3$ the spectrum is suppressed, while for $\bar{f}  <3$ it is enhanced,
 see fig. \ref{plotwave1}.
This behaviour  
can be also deduced from 
 the expression (\ref{Rsr}).  
This approach
 gives informative  analytical expressions and intuitive findings, which we discuss next.

\begin{figure}[t!]
\centering
\includegraphics[width=0.4\textwidth]{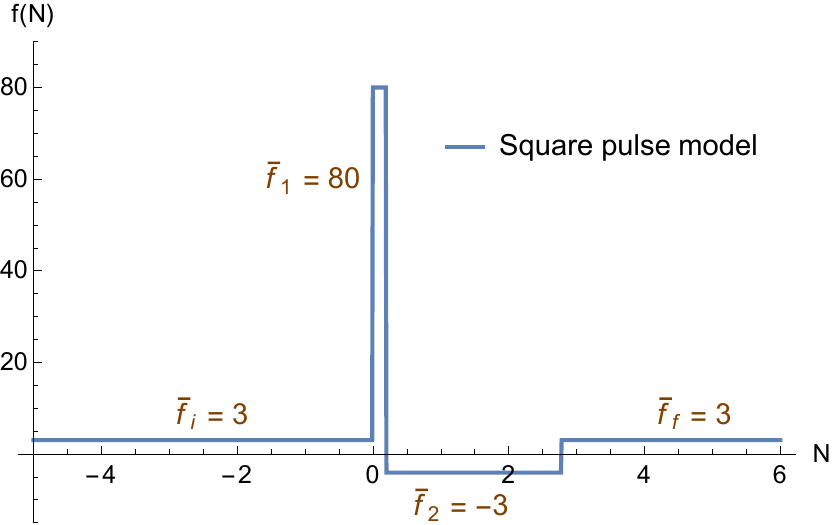}
\hspace{0.5cm} 
 \includegraphics[width=0.4\textwidth]{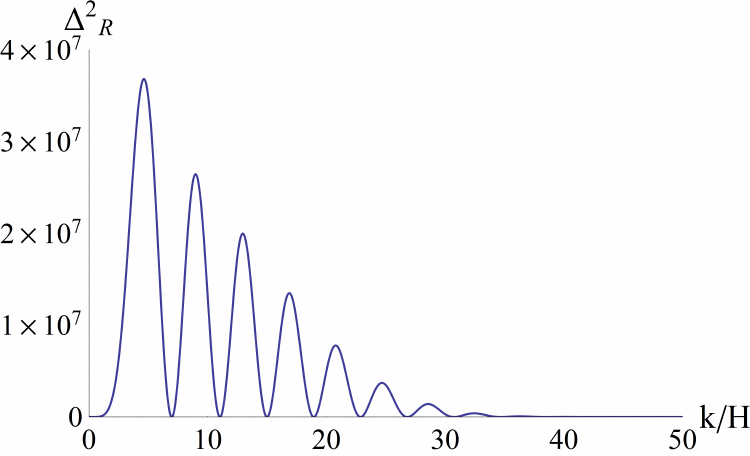} 
\\
\includegraphics[width=0.6\textwidth]{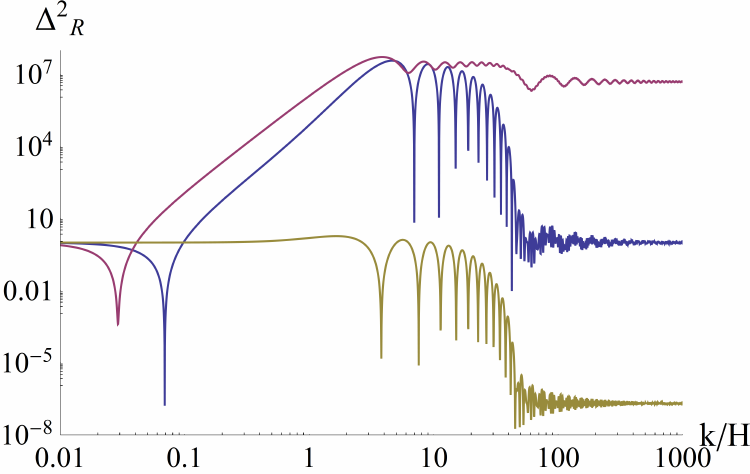}
\caption{
{\it Lower  panel}: A logarithmic plot of the curvature spectrum resulting from a double ``pulse" with  $\bar{f}_1=80$ between 
$N_1=0$ and $N_2=0.2$  and  $\bar{f}_2=-3$ between $N_2$ and $N_3=2.77$ is given by the middle curve depicted with blue color. 
The form of the ``pulses" is shown in the {\it upper left panel}.  The purple upper curve in the main panel is the spectrum
resulting from the negative-friction single ``pulse".  
 The yellow lower curve  is the spectrum resulting from a positive-friction 
single ``pulse". 
The oscillations of the spectrum are also depicted in a linear plot in  the {\it upper right panel}.
}
\label{spectrum1}
\end{figure}

 A {\it first} finding is the size of amplification or the drop  of the power spectrum amplitude.  For the example of a single pulse that lasts $\Delta N=N_2-N_1$ efoldings we find that, at late times $N\gg N_2$ and for large  wavenumbers $k$, the power spectrum is scale invariant and amplified by the factor \cite{Dalianis:2021iig}
\begin{align} \label{enhancement}
\left. \delta \Delta^2_{\cal R} \,\right|_{N\gg N_2}& =|C_m|^2=e^{-(N_{2}-N_{1})(\bar{f}-3)} \approx e^{-\epsilon_2(N_{2}-N_{1})}  \nonumber \\
& =e^{-\int{\epsilon_2 \, dN}}
\end{align}
relative to the CMB amplitude.  
In the second line of 
eq. (\ref{enhancement}) the enhancement is written as an integral after taking the continuum limit.
We  note that analytical expressions for $k \rightarrow 0$ are more difficult to obtain because a component of the matrix $M_{\rm pulse}$ scales as $1/k$ in this limit.
What is of special interest is that the exponent in the above expression is simply the area of the ``pulse" exceeding the value 3.
A general function $f(N)$, that might take positive and negative values,  can be broken in infinitesimal ``pulses", and the enhancement depends on  the integral of $f(N)-3$ over $N$.
For negative $f(N)-3$ there is an increase of power,
with the negative-friction
``pulses" $f(N)<0$  causing a more prominent amplification, whereas  
 positive  $f(N)-3$ causes a decrease.
The positive friction  affects most strongly 
the high-$k$ modes that display a sharp drop to almost zero at a characteristic value of $k/H$.
 For positive and negative ``pulses” with integrated areas that cancel, we find that the low- and high-$k$ modes are not affected by the presence of the feature. As a result the scale-invariant form of the spectrum is modified only for a finite range of wavenumbers $k$, see fig. \ref{spectrum1}.

A {\it second}  finding is the appearance of oscillatory patterns in the amplitude that we discuss in the following.

\subsection{Oscillatory patterns}

The solutions of eq. (\ref{RN}) have an amplitude that gets suppressed or enhanced, depending on the sign of the friction parameter $f(N)$. 
Additionally,  we observe oscillatory patterns 
for particular forms of the friction parameter. 
The origin  of oscillations in the amplitude lies in the 
impact of the strong feature on the phase of the complex quantity ${\cal R}_k$.
It is known that it is possible to obtain an oscillatory pattern in the spectrum if inflation stops for a certain time interval, so that modes that had exited the horizon reenter, or if there is a change in the sound speed \cite{Ballesteros:2018wlw}.  
The analysis with the ``pulse" approximation reveals a  general pattern: 
 An oscillatory spectrum appears  whenever a feature occurs during the evolution of the perturbations that {\it detunes the relative phase between the real and imaginary parts} of the solution ${\cal R}_k$.

A transparent picture of the oscillatory patterns in the spectrum is obtained
in the approximation of square ``pulses" which, though  rough, gives analytic results that make possible the  finding of the characteristic frequencies.
In particular, the oscillatory behaviour of the solutions can be observed in the matrix $M_{\rm pulse}$.
For a single ``pulse" the leading contribution to the power spectrum comes from the $(2,2)$ component \cite{Dalianis:2021iig},
  \begin{equation} \label{Mpulseasympt22}
 ( M_{\rm pulse})_{22}  \approx e^{-\epsilon_2(N_{2}-N_{1})/2}+\epsilon_2\frac{ H }{8\, k} \left[ -2 e^{N_{1}}\sin\left(2 e^{-N_{1}}\frac{k}{H}\right)+2 e^{N_{2}}\sin\left(2 e^{-N_{2}}\frac{k}{H}\right) \right].
  \end{equation}
 It is the subleading corrections  in $H/k$ that introduce oscillatory patterns in the spectrum $|C_m|^2$ with  characteristic periods that can be deduced from eq. (\ref{Mpulseasympt22}). 
The spectrum is expected to vanish at intervals $\delta k/H=e^{N_1}\pi$ and 
$\delta k/H=e^{N_2}\pi$ and  interference patterns appear for $N_1\simeq N_2$.  
For a  sequence of two or more ``pulses" the effect is not simply additive due to the presence of mixing terms. 

An example of square ``pulses" that approximate the friction  function $f(N)$ is depicted in the upper left panel of  fig \ref{spectrum1}. 
Let us discuss it in some detail and compare the exact numerical results displayed  in the plots with our analytical results. 
The feature consists of a positive-friction ``pulse" with $\bar{f}_1 = 80$ in the interval between $N_1 = 0$ and $N_2 = 0.2$, followed by a negative-friction ``pulse" with $\bar{f}_2 = -3$ in the interval between $N_2 = 0.2$ and $N_3 = 2.77$.
In this example we have fine-tuned the integral 
 of $f(N) - 3$ over $N$ to zero.
 The value of the spectrum for a given value of $k/H$ is equal to $|C'_{m}|^2$, where $(C'_{p},C'_{m})$ is given by eq. (\ref{coefficients}) with $M_{\rm pulse}$ being a product of three matrices.
The resulting curvature power spectrum is depicted by the middle blue curve. 
The amplification as well as the characteristic modes and the interference patterns between them are visible.
In the same figure we also display the spectra that would result from a single ``pulse", ignoring the presence of the second. 
The  strong increase of the spectrum results from the effect of the negative-friction ``pulse". The full spectrum for the two ``pulses" has a maximal value that is comparable to that of the negative-friction single-``pulse" spectrum. A rough estimate can be obtained from the asymptotic value of the spectrum for a
single negative ``pulse", which according to eq. (\ref{enhancement}) is $e^{(N_3 - N_2)(3 - \bar{f}_2)} = {\cal O}(10^7)$. 
For the positive-friction single-``pulse" spectrum  a sharp drop to almost zero at a characteristic value of $k/H$ is displayed, as expected.  Positive friction affects most strongly the high-$k$ modes with the Bessel functions
having a zero at a value of their argument roughly equal to $\bar{f}_1/2$.
Also  the spectra display oscillations with characteristic scales
$\delta k/H \approx e^{N_1} \pi = 3.1$, $\delta k/H \approx e^{N_2} \pi = 3.8$ and $\delta k/H \approx  e^{N_3} \pi = 50$,  consistently with our analysis, see  fig. \ref{spectrum1}.

The conclusion drawn  is that strong features in the background evolution can induce a scale-dependent spectrum  which displays, apart from an enhancement that can be  several orders of magnitude large, strong oscillatory patterns. This is clearly visible in the example presented.
These patterns are expected to become less prominent when the features in $f(N)$ become smoother.

\subsection{Step-like transitions versus approximate inflection points  }

For an inflationary model with  an approximate inflection point, 
the parameter  $\epsilon_2$  admits values  $\epsilon_2 \lesssim  -6$. This is realized when
the inflaton crosses a sequence of
a shallow local minimum  and a local  maximum. 
 The inflaton slows down  towards the hilltop but it has enough energy  to cross the local potential well and transit into a second slow-roll phase.
In that region the slope $V'(\phi)$ is negative and $\epsilon_2$ can decrease further producing a very strong enhancement of the curvature power spectrum. 
In the example presented in fig. \ref{f2}  it is apparent that the factor $e^{-\int \epsilon_2 \, dN} $, which gives the enhancement of the power spectrum, is larger in the region of the inflection point than in the region of the step.
A single inflection point, if appropriately tuned (or deformed), suffices to  trigger PBH production and source a detectable stochastic GW background.

For a step-like transition the parameter  $\epsilon_2$  is bounded from below by the value $\epsilon_2 \gtrsim -6$. According to eq. (\ref{enhancement}) the enhancement of the power spectrum is constrained to be smaller than $e^{-\int \epsilon_2\, dN }\approx    e^{\,6\, \delta N}$. The duration of the pulse $\delta N$ is limited by the requirement of graceful exit from the USR phase. Hence, the enhancement of $\Delta^2_{\cal R}$ is usually insufficient in order to produce  a PBH population,  and  the secondary GW background is not sufficiently strong to be observable by the current detectors.  However,  an inflationary model with  multiple steps, or a combination of features can realize a significant enhancement of the scalar power spectrum that will additionally  display a strong oscillatory behaviour, as we stressed above, observable also in the stochastic GW background  (see fig. \ref{figGWrad} of sec. \ref{sGWs}).

\section{Turns without  slow-roll in two-field inflation} \label{sec-turns}

The nature of the inflaton field  remains elusive.
 An inflationary phase realized through more than one fields is an attractive scenario and, moreover, a  viable possibility,
 given that multi-field inflation models 
 make their multi-dimensional dynamics manifest at scales smaller than those directly observed on the CMB sky.
 The phenomenology of interest in multi-field  models is related to the fact that the curvature perturbations can evolve even on super-Hubble scales because of the presence of isocurvature perturbations \cite{Starobinsky:1994mh, Sasaki:1995aw, Garcia-Bellido:1995hsq, Linde:1996gt, 
Langlois:1999dw, Gordon:2000hv, Tsujikawa:2000wc}.

  When more than one fields are  rolling, one can define an adiabatic perturbation component along the direction tangent to the background classical trajectory, and isocurvature perturbation  along the directions orthogonal to the trajectory \cite{Gordon:2000hv}. Curvature perturbations may be affected by the isocurvature perturbations if the background solution follows a curved trajectory in field space. 
In the particular case of a sharp turn  
\cite{Achucarro:2010da, Achucarro:2010jv, Chen:2011zf, Shiu:2011qw, Cespedes:2012hu, Avgoustidis:2012yc, Gao:2012uq, Achucarro:2012fd, Konieczka:2014zja},  
the impact of the isocurvature modes is more prominent, with the resulting curvature spectrum departing significantly from scale invariance. 
A significant amplification takes place if during the sharp turn 
the isocurvature modes experience a transition from heavy to light 
\cite{Palma:2020ejf, Fumagalli:2020adf, Fumagalli:2020nvq,   Fumagalli:2021mpc}.

A turn, even though it is a distinct feature,  admits a similar  description to that of steps at the level of the slow-roll parameters. This becomes apparent if we decompose, along with the perturbations, the second slow-roll parameter into its tangent $\eta_{\parallel}$ and orthogonal  $\eta_{\perp}$ components. Sharp steps lead to large positive values of  
$\eta_{\parallel}$, while sharp turns result in large $|\eta_{\perp}|$ values.
Allowing both $\eta_{\parallel}$ and $|\eta_{\perp}|$ to become large leads to a combination of effects.
When the leading effect comes from the rapid change of the $\eta_{\parallel}$ component, the inflationary dynamics is effectively described by a single-field theory. 
 The $\eta_{\perp}$   component is rapidly changing when a turn in field space takes place. The isocurvature perturbations get excited and through their coupling to the curvature perturbations, which is proportional to $\eta_{\perp}$,    play a crucial role in shaping the final curvature spectrum.

\subsection{Background and perturbations evolution}

We consider an  action for two fields of the form
\be
S=\int d^4x\,\sqrt{-g} \left[\frac{1}{2}R
-\frac{1}{2} g^{\mu\nu}\gamma_{ab} \partial_\mu\phi^a\partial_\nu\phi^b
-V(\phi) \right],
\label{actiontwo} \ee
with $a=1,2$ and unit Planck mass. $\gamma_{ab}$ 
 defines a metric in the internal field space parameterized by the coordinates $\phi^a$.
The Klein-Gordon equation on an expanding, spatially flat background is 
\cite{Achucarro:2012yr, Cespedes:2012hu}
\begin{equation} \label{back1}
\frac{D}{dt}\dot{\phi}^a+3H\dot{\phi}^a+V^a =0\,,
\end{equation}
where $V^a=\gamma^{ab}\partial V/\partial \phi^b$. 
The covariant derivative is defined as $\frac{D}{dt}{\dot{\phi}}^a=\ddot{\phi}^a+\Gamma^{a}_{bc}\,\dot{\phi^b}\,\dot{\phi}^c$. 
The vectors tangent and normal to the path, labeled respectively $T^a$ and $N^a$, are defined  as 
$T^a={\dot{\phi}^a}/{\dot{\phi}}$ and 
$N_a=({\rm det} \gamma)^{1/2}\epsilon_{ab}T^b$.
They satisfy $T^aT_a=N^aN_a=1$, $T^aN_a=0$. 
Projecting eq. (\ref{back1}) along $T^a$, one finds
\be
\ddot{\phi}+3H\dot{\phi}+V_T=0\,,
\label{friedmann} \ee
where $V_T=T^a\partial V/\partial \phi^a$, while $V_N=N^a\partial V/\partial \phi^a$.
The first slow-roll parameter is defined as usual: $\epsilon_1=-{\dot{H}}/{H^2}={\dot{\phi}^2}/(2H^2)$.
The second slow-roll parameter is defined as 
\begin{equation}
\eta^a\equiv -\frac{1}{H\dot{\phi}}\frac{D\dot{\phi}^a}{dt}
\end{equation}
and can be decomposed into the tangent and normal directions, $\eta^a=\eta_\parallel T^a+\eta_\perp N^a$,
where
\begin{equation}
\eta_{\parallel}=-\frac{\ddot{\phi}}{H\dot{\phi}} \quad\quad {\rm and}  \quad\quad \eta_{\perp}=\frac{V_N}{H \dot{\phi}}\,.
\end{equation}

\subsubsection{Perturbations}

The dynamics of the scalar field $\delta \phi$ and   scalar metric $\Psi$ perturbations
can be described by the gauge invariant fields $v^T$ and $v^N$ \cite{Achucarro:2012yr, Cespedes:2012hu},
\begin{equation}
v^T=a T_a \delta \phi^a +a \frac{\dot{\phi}}{H} \Psi\,, \quad  \quad  \quad\quad v^N=aN_a \delta\phi^a\,.
\end{equation} 
A useful parameterization of the perturbations is in terms of curvature and isocurvature fields ${\cal R}$ and ${\cal S}$,
\begin{equation}
{\cal R}=\frac{H}{a\dot{\phi}} v^T \,, \quad\quad \quad 
{\cal S}=\frac{H}{a\dot{\phi}} v^N\,.
\end{equation}
The evolution equations for the curvature  and isocurvature  perturbations 
in two-field inflation,  using the number of efoldings $N$ as independent variable,  can be cast in the form 
\begin{equation} \label{RR}
{\cal R}_{k,NN}+(3+\epsilon_1-2\eta_{\parallel}){\cal R}_{k,N}+
\frac{k^2}{H^2}e^{-2N}{\cal R}_k \approx -2\left( 
\eta_\perp{\cal S}_{k,N}+
\eta_{\perp,N}{\cal S}_k+3\eta_\perp {\cal S}_k \right)
\end{equation}
\begin{equation} \label{SS}
{\cal S}_{k,NN}+ (3-\epsilon_1){\cal S}_{k,N}+
\frac{k^2}{H^2}e^{-2N}{\cal S}_k
+\left(\frac{M^2}{H^2} -\eta_\perp^2 + \epsilon_1\,{\mathbb R}\right){\cal S}_k \approx 
2\, \eta_\perp {\cal R}_{k,N}\,,
\end{equation}
where $\eta_\parallel=\epsilon_1-{\epsilon_{1,N}}/{2\epsilon_1}$.
In the above system of approximate equations we have assumed for simplicity  that $\epsilon_1$ is 
small and roughly constant.
${\mathbb R}$ 
is the Ricci scalar of the  
internal manifold spanned by the scalar fields. 
The Ricci scalar vanishes for a model with standard kinetic terms for the two fields, which is the case of interest to us here. 
The coefficient in front of ${\cal R}_{k,N}$ 
is the friction term $f(N)=3+\epsilon_{2 \parallel} -\epsilon_1$.
The mass $M$ of the isocurvature perturbation is
given by the curvature of the potential in the direction perpendicular to the
trajectory of the background inflaton. 

\subsection{Enhancement}

The friction term is critical for the enhancement of the curvature power spectrum in single field inflation. In the multi-field case the source of amplification 
can be found in the isocurvature perturbations.
We stress that we do not
consider the possibility of a curved field manifold, but assume that the 
fields have standard kinetic terms and we can set ${\mathbb R}=0$.
 We are interested in strong deviations from the slow-roll regime during short intervals
in $N$, 
during which the second slow-roll  parameter
can grow large. 
This happens when  the evolution of the
background inflaton is accelerated  either in the direction of the flow through $\eta_\parallel(N)$, or perpendicularly to it through $\eta_\perp(N)$.
The result is a   
deviation of the spectrum from scale invariance with 
a strong enhancement of the curvature perturbations over a range of momentum scales. 
We list these two main scenaria:
\begin{itemize}
\item
Suppressed isocurvature modes.
This case is implemented for $\eta_\perp^2 \ll M^2/H^2$.  Hence, 
the rhs of eq. (\ref{RR}) vanishes
and the system is effectively described by the dynamics of the single field inflation.
 The curvature mode can be enhanced due to the friction term $f(N)$
that requires large values of the parameter $\eta_\parallel\approx -\epsilon_2/2$. Such values can
be attained if the inflaton potential
displays an inflection point, or a sharp step \cite{Kefala:2020xsx, Dalianis:2021iig}.
This case corresponds to  acceleration in the evolution of the
background inflaton in the direction of the flow.

\item
Excited isocurvature modes.
For $\eta_\perp^2 \gg M^2/H^2$  the isocurvature contribution 
in the rhs of eq. (\ref{RR})  becomes
large and acts as a source for the curvature perturbations, leading to their strong
enhancement. 
We assume 
that this strong feature is effective within a short period.
At a later time $\eta_\perp$ becomes small and the isocurvature
perturbations become suppressed again. 
This case corresponds to  acceleration in the evolution of the
background inflaton in the direction  perpendicular to the flow.
\end{itemize}

In the first case, the inflaton potential must feature either  a near  inflection point or steps   finely-tuned with high accuracy in order to achieve strong enhancement.
In the second case a single sharp turn in the 
inflaton trajectory must reach a value of $4\pi$ for the enhanced spectrum to 
have observable consequences \cite{Palma:2020ejf}. This is possible only if the 
field manifold is curved, i.e ${\mathbb R}\neq 0$.  
For an inflationary set up with canonical kinetic terms an enhancement can be realized if 
a sequence of turns in field space 
occurs within a small number of efoldings  \cite{Boutivas:2022qtl}. 
Prominent  oscillatory patterns appear as well.

\subsection{Describing turns in field space}
  
We consider models of a two-component 
field $\phi^a=(\chi,\psi)$ with vanishing
  field-manifold curvature.
We assume that the potential has an almost flat direction along a curve
$\psi=g(\chi)$ with a small inflationary constant slope  along this direction 
and large curvature along the perpendicular direction, so that the flat direction forms a valley. 
 A particular realization of
this setup, with $g(\chi)=a^2/\chi$, is given in ref. \cite{Cespedes:2012hu}.
The unit vectors, tangential and normal to the valley at
$\phi^a=\left(\chi, g(\chi) \right)$, are
\begin{equation} \label{tangunit}
T^a=\frac{1}{\sqrt{1+g'^2(\chi)}} \left(1, g'(\chi) \right)\,, \quad\quad \quad
 N^a=\frac{1}{\sqrt{1+g'^2(\chi)}} \left(g'(\chi),-1 \right).
\end{equation}
From the relation $\frac{DT^a}{dt}=-H \eta_{\perp} N^a$
we deduce the following expression for the component of the $\eta$ parameter in the normal direction
\begin{equation}
\eta_\perp  \, = \, \frac{g''(\chi)}{1+g'^2(\chi)}\frac{d\chi}{dN}
=\pm \sqrt{2\epsilon_1} \frac{g''(\chi)}{(1+g'^2(\chi))^{3/2}} \, .
\label{etaperp} 
\end{equation}
Regions in which
$g''(\chi)\not=0$ 
 yield  nonzero values for $\eta_\perp$.
Practically, the sharp turn is the intersection of  two linear parts of the valley causing $\eta_\perp$ to 
rise and fall sharply from zero. 
 We place the turn  near $\chi=\psi=0$  
 without loss of generality. 
The angle of rotation in field space is found from the relation 
$d\theta/dN= \eta_\perp$,
and is given by
\begin{equation}
 \Delta \theta =\int_{N_i}^{N_f} \eta_\perp(\chi) dN
=\int_{\chi_i}^{\chi_f} \frac{g''(\chi)}{1+g'^2(\chi)} d\chi=
\Bigl. \arctan\left( g'(\chi) \right) \Bigr|_{\chi_i}^{\chi_f}.
\label{thetatotal} 
\end{equation}
The maximal angle is $\Delta\theta=\pi$ and can be obtained for $g'(\chi_i)\to-\infty$ before and $g'(\chi_f)\to \infty$ after the turn.
The fact that the value of the integral 
is bounded by $\pi$ means that the effect of a sharp turn on the amplification of 
the isocurvature and curvature perturbations is limited.  This conclusion changes if multiple turns occur along the flat direction. 
A sequence of turns with alternating signs can produce   a combined effect that can enhance the power spectrum substantially.
An analytic  quantitative description  is possible by  approximating the various features through ``pulses"  \cite{Boutivas:2022qtl}, in a similar fashion with the analysis done for the case of steps in  the previous section.

\subsection{The quantitative features of the evolution} \label{qualitative}

We assume that initially the mass of the  isocurvature modes $M$ is constant and large
($M/H\gg 1$), so that these modes  get suppressed
during the linear parts of the inflationary trajectory.
In the region of the turn  we have $\eta_\perp \gg M/H$  and both the curvature and isocurvature modes
get excited. The system is strongly coupled and the evolution becomes
non-trivial. The term $\sim \eta^2_\perp$ in the lhs of 
eq. (\ref{SS}) acts as a negative mass term, triggering the 
rapid growth of the field ${\cal S}_k$,  which subsequently  sources through eq. (\ref{RR})
the curvature perturbation ${\cal R}_k$. The resulting growth 
of ${\cal R}_k$ backreacts, generating a source term in the rhs of eq. (\ref{SS}) and thus
moderating the maximal value of the field ${\cal S}_k$. 
The combined effect results in the enhancement of both modes. 
After the turn,  $\eta_\perp$ becomes negligible and ${\cal S}_k$ attains 
a nonzero mass that makes it vanish eventually. 
During  the same period the curvature
mode ${\cal R}_k$, which has been enhanced  within a certain momentum range,
freezes.  
In addition to the enhancement, sharp turns, quite similarly to the case of sharp steps, cause the appearance of distinctive oscillations in the curvature spectrum.

We consider again the ``pulse" approximation of the previous section, in which 
$\eta_\perp$ has the form
\begin{equation} \label{etaperppulse}
\eta_\perp(N)=\eta_{\perp 0}\left( \Theta(N-N_1)-\Theta(N-N_2)\right).
\end{equation}
The 
parameter $\eta_\perp$ takes a constant value for $N_1<N<N_2$, and approaches zero very
quickly outside this range.
The total area of 
the ``pulse" corresponds to the total angle of the turn.
During the linear part of the trajectory, i.e  $N<N_1$ and $N>N_2$, 
the two equations of motion, (\ref{RR}) and (\ref{SS}), decouple
\begin{eqnarray}
\mathcal{R}_{k,NN}+3\,\mathcal{R}_{k,N}+\frac{k^2}{H^2}e^{-2N}\mathcal{R}_k&=&0
\label{eqmot1} \\
\mathcal{S}_{k,NN}+3\,\mathcal{S}_{k,N}+\frac{k^2}{H^2}e^{-2N}\mathcal{S}_k+\frac{M^2}{H^2}
\mathcal{S}_k &=&0,
\label{eqmot2}
 \end{eqnarray}
and the solutions are a combination of Bessel functions:
\begin{eqnarray}
\mathcal{R}_k(N)&=&e^{-3N/2}\left[C_pJ_{\frac{3}{2}}\left(e^{-N}\frac{k}{H}\right)+C_mJ_{-\frac{3}{2}}\left(e^{-N}\frac{k}{H}\right)\right]
\label{eqmotsol1} \\
\mathcal{S}_k(N)&=&e^{-3N/2}\left[D_pJ_{\frac{1}{2}\sqrt{9-4M^2/H^2}}
\left(e^{-N}\frac{k}{H}\right)+D_mJ_{-\frac{1}{2}\sqrt{9-4M^2/H^2}}
\left(e^{-N}\frac{k}{H}\right)\right].
\label{eqmotsol2} \end{eqnarray}
Initial conditions corresponding to the
Bunch-Davies vacuum are obtained for $C_{p}=1$, $C_{m}=i$.
For the massive mode ${\cal S}_k$ the coefficients are chosen so that they
reproduce the free-wave solution at the initial stage of inflation.
They have the form $D_p=-\sqrt{2}(1+i) \, f_1(M, H, \varphi)$, $D_m=\sqrt{2}(1+i) \, f_2(M, H, \varphi)$ where the functions $f_1$ and $f_2$ can be found in \cite{Boutivas:2022qtl}.

The evolution in the interval $N_1<N<N_2$ is complicated because of the coupling 
between the two modes. 
The duration of the interval  $\Delta N=N_2-N_1\lesssim 1$ is short enough
so that  it is a good approximation to neglect the expansion of space and simplify 
the evolution equations:
\begin{eqnarray}
\mathcal{R}_{k,NN}+\frac{k^2}{H^2}\mathcal{R}_k+2\eta_{\perp 0}\mathcal{S}_{k,N}&=&0
\label{eveqsimple1} \\
\mathcal{S}_{k,NN}+\left(\frac{k^2}{H^2}+\frac{M^2}{H^2}-\eta_{\perp 0}^2\right)\mathcal{S}_k
-2\eta_{\perp 0}\mathcal{R}_{k,N}&=&0.
\label{eveqsimple2} \end{eqnarray}
Following refs. \cite{Palma:2020ejf, Fumagalli:2020nvq}, we look for solutions of the form 
\be
\mathcal{R}_k=A e^{\omega N}, 
\;\;\;\; 
\mathcal{S}_k=B e^{\omega N}.
\label{ABi} \ee
There are four independent solutions $\omega_1$, $\omega_2$,  $\omega_3$ and  $\omega_4$. 
 The following relations between $A_i$ and $B_i$ must hold
 \begin{equation} \label{AB}
 B_{1,2}=h_+\, \omega_{1,2}\, A_{1,2} \,, \quad\quad\quad
 B_{3,4}=h_-\, \omega_{3,4}\, A_{3,4},
 \end{equation}
 where $h_+$ and $h_-$ are functions of $\eta_{\perp 0}$, $M/H$ and $k/H$ \cite{Boutivas:2022qtl}.
The solutions on either side of $N_1$ and $N_2$ can be matched, assuming the
continuity of ${\cal R}_k(N)$, ${\cal S}_k(N)$ and ${\cal S}_{k,N}(N)$. 
The first derivative of ${\cal R}_k(N)$ must account for
the $\delta$-function arising from the derivative of $\eta_\perp(N)$ at these 
points. This leads to the conditions
\begin{eqnarray}
\mathcal{R}_{k,N}(N_{1-})&=&\mathcal{R}_{k,N}(N_{1+})+2\eta_{\perp 0}\mathcal{S}_k(N_1)
\nonumber \\
\mathcal{R}_{k,N}(N_{2-})&=&\mathcal{R}_{k,N}(N_{2+})-2\eta_{\perp 0}\mathcal{S}_k(N_2).
\label{Rder1} \end{eqnarray}

The  values of the coefficients  $C_{p},C_{m},D_{p},D_{m}$ before 
the ``pulse" , see eq. (\ref{eqmotsol1}),
 are determined by the initial conditions. 
The coefficients  $A_i$ 
within the ``pulse"  (\ref{AB}) and eventually  the final coefficients of the free solutions 
$C'_{p},C'_{m},D'_{p},D'_{m}$ after 
the ``pulse"
 can be found through the use of the boundary conditions (\ref{Rder1}).
In analogy to the previous section, one 
can  determine the matrix $M_{\text{pulse}}$ that links the
solutions before and after the ``pulse":
\be
\begin{bmatrix}C'_{p}\\C'_{m}\\D'_{p}\\D'_{m}\end{bmatrix}
=M_{\text{pulse}}(N_1,N_2,k,M,\eta_\perp)
\begin{bmatrix}C_{p}\\C_{m}\\D_{p}\\D_{m}\end{bmatrix}.
\label{matrix44} \ee
The  elements of the matrix $M_{\rm pulse}$ are  lengthy and we do not present them here explicitly. 
This matrix  facilitates calculations for more general problems with multiple ``pulses",
occurring when 
the linear valley of the potential is interrupted by multiple, 
successive turns.

\begin{figure}[t!]
\centering
 \includegraphics[width=0.48\textwidth]{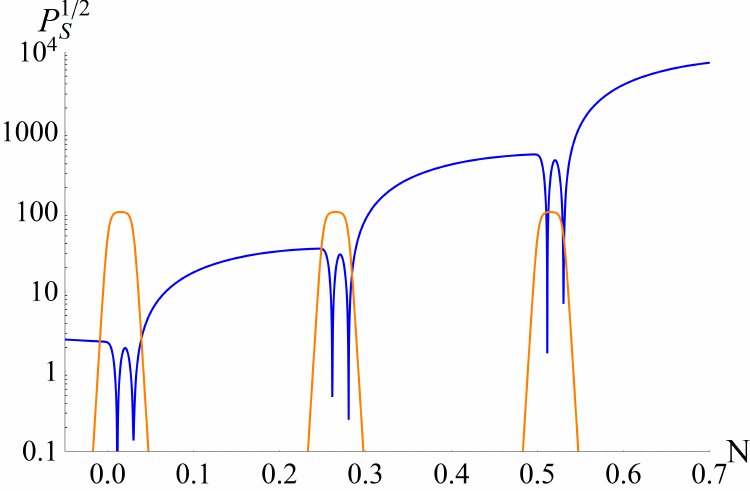}
\includegraphics[width=0.48\textwidth]{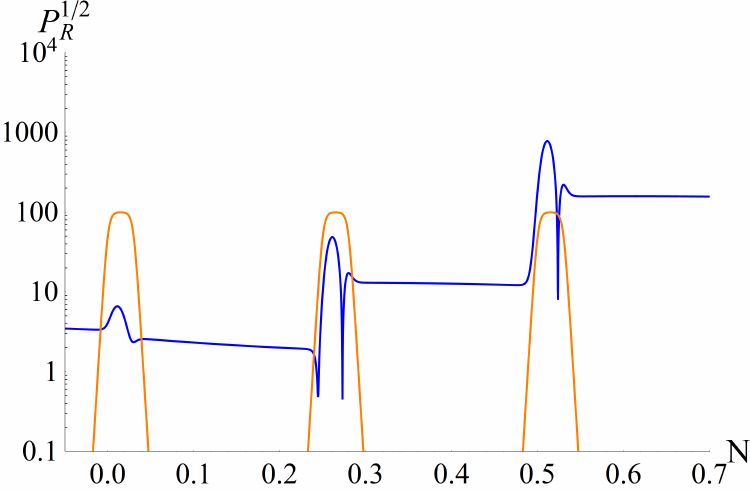}
\includegraphics[width=0.48\textwidth]{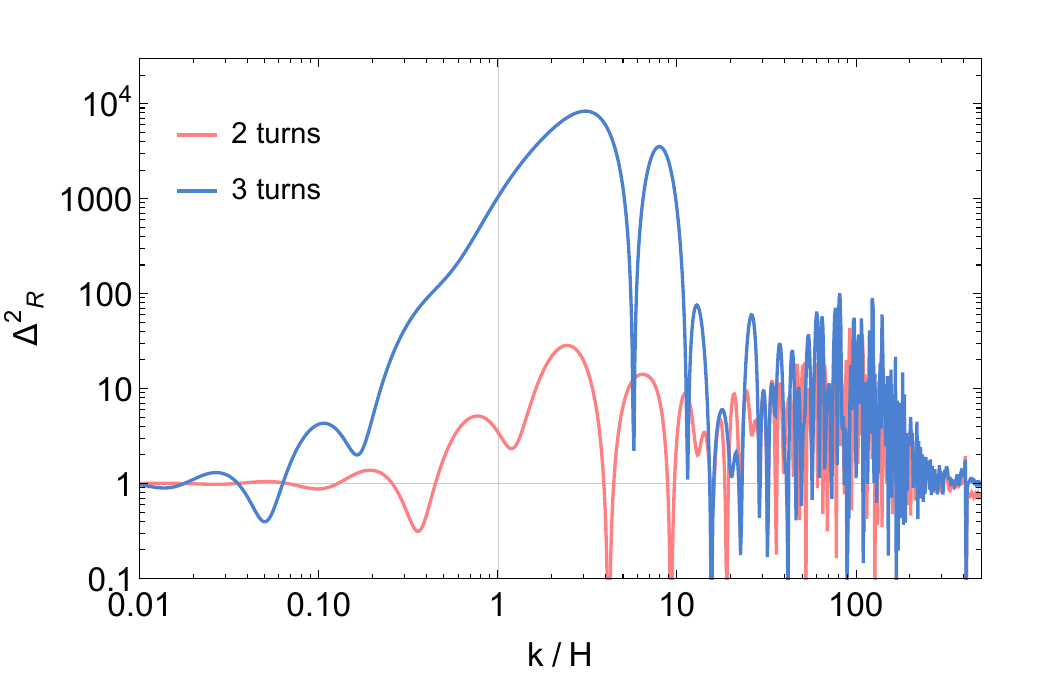}  \includegraphics[width=0.48\textwidth]{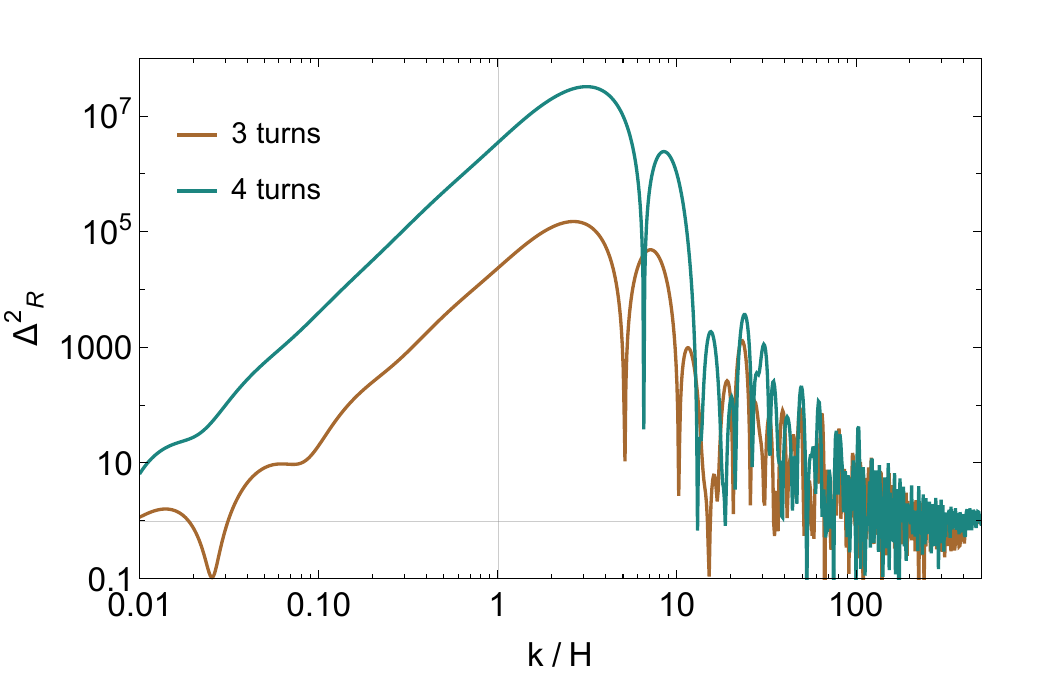}
\caption{
{\it Upper panels}: 
The evolution of the isocurvature mode (left plot) 
and curvature mode (right plot) are  depicted in blue curves,  
for a time dependent parameter $\eta_\perp(N)$ associated with three sharp turns.
The function $|\eta_\perp(N)|$ is
 depicted 
in orange curves
as  a sequence of three ``pulses" of alternating sign. 
{\it Lower panels}:
The corresponding curvature spectrum  
is depicted in blue (left panel).
In pink color  we depict the curvature spectrum produced if we keep two out of the three turns. 
The amplitude of the  curvature spectrum is enhanced when we increase the sharpness of the pulses, as was done 
for three and four turns (right panel).
}
\label{figPSPR}
\end{figure}

The impact of the ``pulse" on the evolution of the perturbations can be inferred
through inspection of the solutions for $\omega$. 
The solutions $\omega_{3,4}$ are purely imaginary, resulting in 
oscillatory behaviour. 
The other two  $\omega_{1,2}$ can be either real or imaginary
depending on the parameters of the problem. 
For $\frac{k}{H}\geq\sqrt{\eta_{\perp 0}^2-\frac{M^2}{H^2}}$ 
they are imaginary as well, but for $\frac{k}{H}\leq\sqrt{\eta_{\perp 0}^2-\frac{M^2}{H^2}}$
they become real. It is the real solution that induces an exponential growth or suppression of the curvature power spectrum. 
In the limit $\Delta N\rightarrow 0$, $\eta_{\perp 0}\to\infty$, with  the total angle of the turn $\Delta\theta=\eta_{\perp 0} \Delta N$ kept constant, we have the real solutions
\be
\omega_{1,2}=\pm \frac{k}{H\sqrt{3}}+\mathcal{O}(\Delta N).
\label{infi}\ee
For turns that yield real $\omega_{1,2}$  
the spectrum is enhanced by a factor \cite{Boutivas:2022qtl}
\begin{equation}
\delta \Delta^2_{\cal R} \sim \exp{\left[\frac{2k}{H\sqrt{3}}\Delta N\right]}= \exp{\left[\frac{2\Delta\theta}{\sqrt{3}}\frac{k/H}{\eta_{\perp 0}}\right]}.
\label{estenh} 
\end{equation} 
For inflation models with canonical kinetic terms  
where  $\Delta\theta \lesssim \pi$ 
the expression (\ref{estenh})  
 implies that 
the enhancement of the spectrum appears for scales 
$k/H$ comparable to $\eta_{\perp 0}$. 
The general solution is a superposition of all independent solutions (\ref{ABi}), 
 hence the exponential growth is accompanied by oscillatory behaviour with
a characteristic frequency set by the height of the ``pulse" $\eta_{\perp 0}$.

The constraint on $\Delta \theta$ implies that a single turn results only in
limited growth of the spectrum \cite{Palma:2020ejf}. However, multiple turns can 
have an additive, and in particular conditions resonant, effect. In this respect it is important to
point out another feature of the solutions. 
We are interested in the regime $\eta_{\perp 0}\gg M/H$, so that we can neglect the
effect of the mass inside the ``pulse". 
For $k/H\ll \eta_{\perp 0}$ the solution $\omega_{1,2}$ is imaginary and  strong oscillations, with a frequency set by $\eta_{\perp 0}$, occur
within the ``pulse". Outside the ``pulse"
the characteristic frequency of oscillations
is set by $k$ or $M$ and is also modulated by the expansion. 
The continuity of ${\cal S}_k(N)$ and its derivative implies that
the amplitude of oscillations increases significantly when the perturbation 
exits the ``pulse", see fig. \ref{figPSPR}.

In fig. \ref{figPSPR} we present some specific examples
of  inflationary set-ups with  two fields and vanishing 
internal curvature ${\mathbb R}$ that exhibit 
 trajectories with sharp turns.
 In the top row
 the evolution of  ${\cal S}_k(N)$ (left plot) and 
 ${\cal R}_k(N)$ (right plot) is depicted  for three “pulses” corresponding to three sharp turns of alternating direction. The pulses are narrow and each turn $|\Delta \theta|$ in field space is smaller than $\pi$.
 The mass of the isocurvature mode is  $M/H \approx 5$.  
 The distinctive feature is the strong increase of the amplitude of the isocurvature mode when the perturbation exits the ``pulse”.  
The first ``pulse" is at the time that we have set $N = 0$ for the number of efoldings and 
the  ``pulses"  are repeated with a sequence such that a resonance effect occurs, with the amplitude of ${\cal S}_k$ being amplified each time a “pulse” is traversed. This effect triggers the growth of the curvature mode.

\begin{figure}[t!]
\centering
\includegraphics[width=0.47\textwidth]{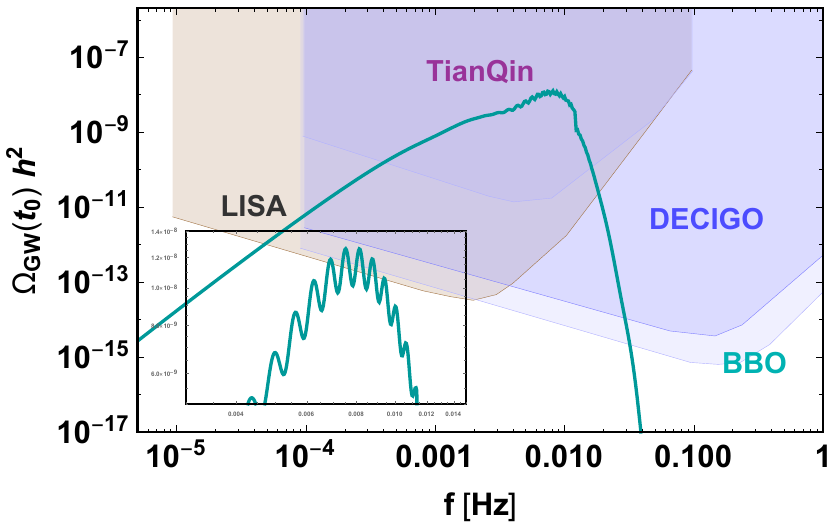}
\hspace{0.5cm}
 \includegraphics[width=0.47\textwidth]{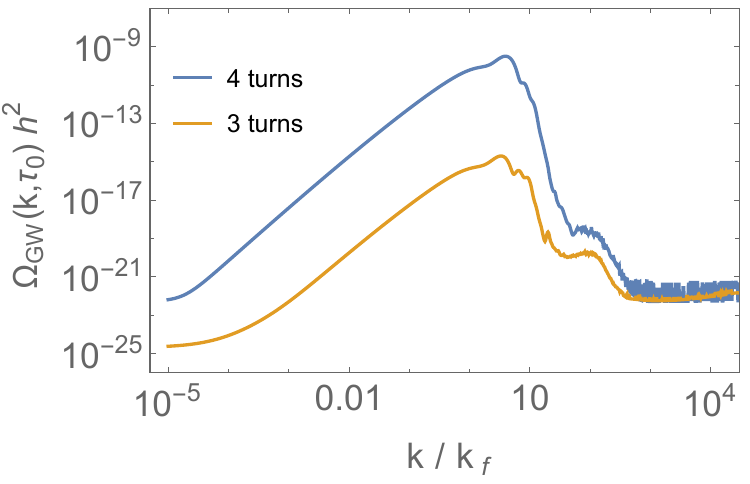} 
\caption{
{\it Left panel}:  The  GW density parameter produced by  an inflationary model that has two strong  features:  a step and an inflection. A zoom in about the peak is also depicted.
{\it Right panel}: The GW spectral density produced by inflationary trajectories with 3 and 4 successive turns. The scale $k_f$ is the scale that crosses the horizon at the time when the first feature occurs.
}
\label{figGWrad}
\end{figure}

In the bottom row of fig. \ref{figPSPR}  the curvature power spectrum  induced by the aforementioned  ${\eta_{\perp}(N)}$ is shown.  
Also, the  curvature spectra induced by two, three and four turns are displayed. 
 The spectra are normalized to the scale-invariant one and the location of the point $k/H = 1$ is arbitrary.
The spectra in the right panel are generated by rather sharp turns and produce an amplified spectrum $\Delta_{\cal R}^2$ that is possibly testable via the induced stochastic GW background (see fig. \ref{figGWrad} of sec. \ref{sGWs}).

\section{Inflationary models with special features}

Embedding models of inflation with strong features  in a high energy physics framework is a rather challenging task.  
 A fundamental quantum field theory
can be predictive and can reveal  aspects of the high energy scale related to the strong feature. In the following we will describe four models: one with step-like transitions and three with an approximate inflection point.
We will not present here a model with a sharp turn. A concrete model can be found e.g. in ref. \cite{Bhattacharya:2022fze}. 

\subsection{Step-like features in the inflaton potential }

The steps connect regions in which the potential is nearly flat.
The basic pattern corresponds to the vacuum energy having one or more transition points at which it jumps from one constant value to another.
One can speculate that these points may correspond to 
filed values  at which 
certain modes decouple very quickly. 
If these modes have quantum fluctuations that  contribute 
to the vacuum energy a step-like transition may be realized.
Decoupling effects become visible when the effective 
potential is regularized in a mass-sensitive scheme.
The renormalization-group equation for the potential
can capture the decoupling behavior.  
However, a concrete quantitative calculation of these effects is elusive due to the  fundamental lack of understanding of the
nature of vacuum energy or the cosmological constant.

Some intuition  can be obtained by considering 
the role of underlying symmetries. A specific framework, which can be used as a basis, is provided by the $\alpha$-attractors in supergravity \cite{Kallosh:2013hoa, Ferrara:2013rsa}.
The theory contains  two fields and has conformal invariance and a global 
${\rm SO}(1,1)$ symmetry.
After one of the fields is eliminating through gauge fixing, 
the following Lagrangian is obtained
\begin{equation}
{\cal L}=\sqrt{-g}\left[ 
 \frac{1}{2} R(g)-\frac{1}{2}\partial_\mu \phi \partial^\mu \phi
-F^2\left(\tanh \frac{\phi}{\sqrt{6 \alpha}} \right) \right]\,,
\label{lagralphagauge} 
\end{equation}
where an additional free parameter $\alpha$ has been included.
A constant function $F(x)$ preserves the ${\rm SO}(1,1)$ symmetry of the initial lagrangian,  
and corresponds to a cosmological constant in the gauge-fixed version.
 The value of the cosmological constant is not constrained by the symmetry and is arbitrary.
A minimal deformation of the symmetry takes place  by assuming that $F(x)$ 
takes different values 
over  successive ranges of $x$, with a rapid transition in between. 
More generally, we can assume that 
$F(x)$ has the schematic form
\be
F(x) \sim x^n+ \sum_i A_i\, \Theta(x-x_i),
\label{fx} \ee
where  $\Theta$ is a step function. 
It is understood that  each step-function is replaced by a continuous function with a sharp transition
at $x_i$ so that the  potential varies smoothly.
 In the framework of 
 $\alpha$-attractors the dependence of the potential on
 $\tanh ({\phi}/\sqrt{6\alpha})$, with $\alpha$ a free parameter, allows for
 potentials with transitions of arbitrary steepness, see the right panel of fig. \ref{graphs}. 
This is an example where the function $F(x)$ gives
 potentials which are
generalizations of the Starobinsky model 
 \cite{Starobinsky:1980te}, with the addition of one or more steep steps.

\subsection{Approximate inflection-point inflationary models}

A primordial scalar spectrum with a maximal enhancement can be realized  through inflationary potentials that contain an approximate inflection point. This sort of models feature a shallow potential well (local maximum) that separates potential slopes where  the initial and the final slow-roll phases are realized.
We will briefly present the  models of refs. \cite{Dalianis:2018frf, Nanopoulos:2020nnh}.

\subsubsection{$\alpha$-attractrors}

In the context of supergravity and superconformal theory,  the general class of $\alpha$-attractor models  have the advantage of flattening the scalar potential for large field values  while exhibiting an inflection point for small field values \cite{Dalianis:2018frf}. 
The necessary conditions the superpotential function $F(x)$ of the scalar Lagrangian (\ref{lagralphagauge}) should satisfy for the appearence of an approximate inflection point are $V'(\phi_{\rm infl})\approx 0$ and  $V''(\phi_{\rm infl})=0$.
For particular choices of the parameters,  the model can be compatible with the CMB constraints \cite{Dalianis:2018frf, Dalianis:2019asr, Iacconi:2021ltm}.
 
 For a polynomial $F(x)$ the simplest choice  with one  inflection point is the cubic.
The corresponding inflationary trajectory for a canonically normalized field is determined by the potential 
\begin{equation}\label{dims}
V=V_0 \left[c_0+
c_1\tanh\left(\frac{\phi}{\sqrt{6\alpha}}\right)+c_2\tanh^2\left(\frac{\phi}{\sqrt{6\alpha}}\right)+c_3\tanh^3\left(\frac{\phi}{\sqrt{6\alpha}}\right)\right]^2 \,,
\end{equation}
where $c_0$, $c_1$, $c_2 $ and $c_3$ are constants, depicted in the left panel of fig. \ref{graphs}.
Another choice is   $F(x)\sim \lambda\Big(x+A\sin(x)\Big)$
 and the potential turns out to be
\begin{eqnarray}\label{chaomod}
V=\lambda^2  \Big[\tanh(\phi/\sqrt{6})+A\sin\left(\tanh(\phi/\sqrt{6})/f_\phi\right)\Big]^2\,,
\end{eqnarray}
having taken $\alpha=1$ and $A$, $\lambda$, $f_\phi$ constants.
Another example in which a proper inflection point can be generated is found in ref. \cite{Dalianis:2019asr}  where the function $F(x)$ is built in terms of exponential functions.

PBH formation  requires particular values for the parameters  of the above potentials and these are determined via a subtle numerical process.
Firstly, a central PBH mass  has to be chosen 
and this determines the $k$-position for the $\Delta^2_{\cal R}(k)$ peak. Secondly, the PBH probability function  is exponentially sensitive to the   amplitude of the 
 peak and its precise value is found after a delicate  selection of the potential parameters.
 The spectrum $\Delta^2_{\cal R}(k)$ is reliably found only by solving numerically the Mukhanov-Sasaki equation (\ref{eqMS}). 
It goes without saying that consistency with the CMB normalization, the measured $n_s$ and $\alpha_s$ values, as well as adequate efoldings $N_*$ have to be reproduced.

\subsubsection{No-scale supergravity}

A natural framework for formulating
models of inflation is supergravity. 
Potentials that depend on a minimal number of parameters and  evade the so-called $\eta$ problem 
can be found in no-scale supergravity models \cite{Lahanas:1986uc}. These  also emerge  as the low energy limit of compactified string models.
 No-scale models are necessarily multi-field models, and apart from the inflaton, they include additional moduli scalar fields.
An inflationary model based on no-scale supergravity \cite{Ellis:2013xoa} can yield  a Starobinsky-like potential. 
By  deforming the ordinary ${\rm SU}(2,1)/ {\rm SU}(2) \times {\rm U}(1)$ K\"ahler potential an inflection point 
can  appear.
This is achieved by introducing
an exponential term with one extra parameter. 
All the phenomenological constraints at the CMB scales  can be satisfied, also producing 
PBHs with a significant cosmological population.

\begin{figure}[t!]
\centering
\includegraphics[width=0.52\textwidth]{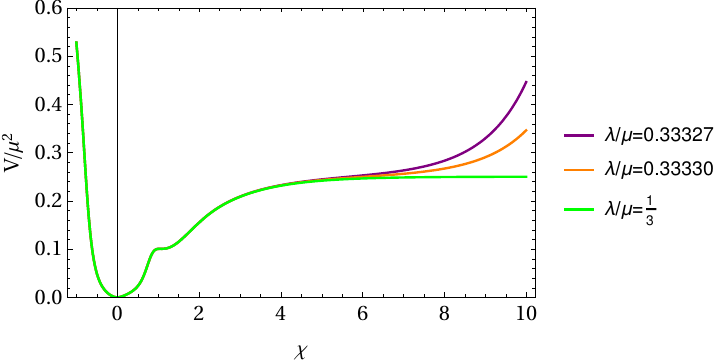}
\hspace{0.4cm}
 \includegraphics[width=0.38\textwidth]{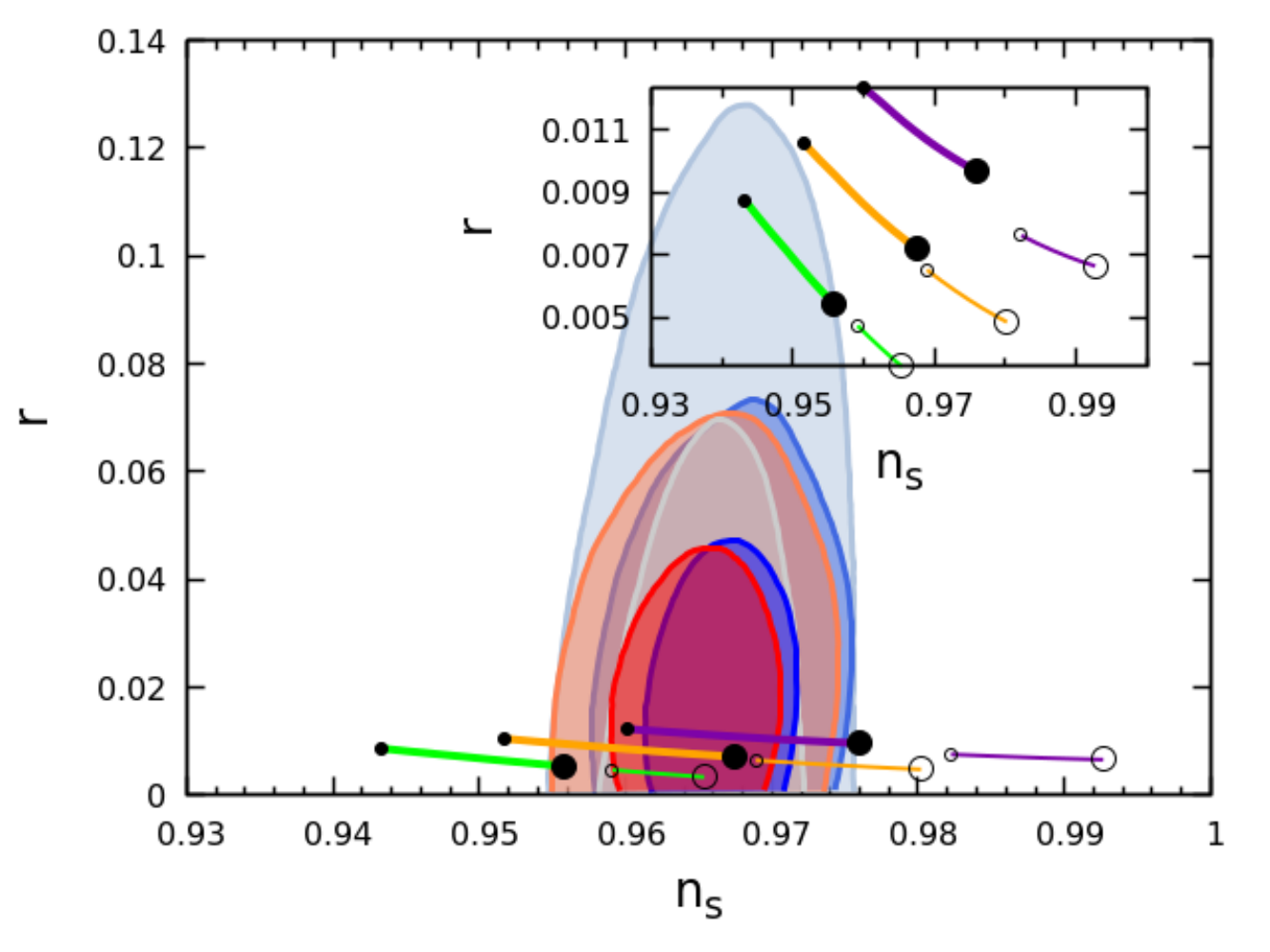} 
\caption{
{\it Left panel}:   The potential given in eq. (\ref{nanop}) as a function of $\chi$, for various values of the ratio $\lambda /\mu$.
{\it Right panel}: 
The predictions for the tilt $n_s$ and the tensor-to-scalar ratio $r$ \cite{Nanopoulos:2020nnh}.}
\label{figspan}
\end{figure}

Modifications of the K\"ahler potential that induce an inflection point to the scalar potential have the form
\begin{equation}
K=-3 \ln \left[ T+\bar{T} -\frac{\varphi \, \bar{\varphi}}{3} + a\, e^{-b(\varphi +\bar{\varphi})^2}\left(\varphi+ \bar{\varphi} \right)^4\right],
\end{equation}
where $T$ and $\varphi$ are two chiral superfields that parametrize the noncompact  ${\rm SU}(2,1)/{\rm SU}(2) \times {\rm U}(1)$  coset space, and $a$,  $b$ are real numbers.
The real part of the field $\varphi$ plays the role of the inflaton. The simplest superpotential is the Wess-Zumino model, $W=\mu \varphi^2/2 -\lambda \varphi^3/3$, 
characterized by a mass term $\mu$ and a trilinear coupling $\lambda$. The scalar potential along the direction $T=\bar{T}=c/2$, $\text{Im} \, \varphi=0$, is
\begin{equation} \label{nanop}
V(\phi)=\frac{3 e^{3b \phi^2 (c\mu^2- 2\sqrt{\lambda} \mu \phi +3 \lambda^2 \phi^2)}}{[-3a \phi^4 +e^{b\phi^2}(-3c +\phi^2)]^2 \,[e^{b \phi^2} -6 a \phi^2 (6 +b\phi^2(-9+2b\phi^2))]}\,,
\end{equation}
where $\phi=\text{Re} \,\varphi$ and $c$ a constant. 
After a field transformation that puts the kinetic term in canonical form, $\frac12 \partial_\mu \chi \partial^\mu \chi=K_{\phi \phi} \partial_\mu \phi \partial^\mu \phi $, the above potential  can be expressed in terms of the $\chi$ field, see fig. \ref{figspan}.

One can notice that for particular values of the parameters the potential has the required features which ensure that a sizable abundance of PBH is created along with induced GWs \cite{Spanos:2022euu}.
This is achieved via  the approximate inflection point, which results   from a fine-tuning of the parameters  in the modification of the K\"ahler potential.

\section{Secondary GWs} \label{sGWs}

 The GW astronomy has started exploring a new range of cosmological phenomena via a network of operating GW detectors. From the cosmological perspective the detection of a cosmic gravitational radiation background would be a milestone in our quest to understand the universe. 
Stochastic GW backgrounds are thought to originate in several physical sources  \cite{Kuroyanagi:2018csn, Christensen:2018iqi, Caprini:2018mtu}.  
Among them there is a unambiguous source: the primordial scalar perturbations that source tensor modes at the non-linear level of the cosmological  perturbation theory \cite{Matarrese:1992rp, Matarrese:1997ay, Noh:2004bc, Carbone:2004iv, Ananda:2006af, Baumann:2007zm}, the so-called secondary or induced gravitational waves (IGWs).

Once the primordial density perturbations enter the horizon, during the adiabatic expansion of the universe, they start  interacting and evolving.
The non-linear combinations that act as tensors have an amplitude given roughly  by the square of the curvature perturbations.  
Induced tensors are suppressed 
by the observed small value of  $\delta\rho/\rho $ at the CMB scales. 
Nevertheless,  IGWs  can be sizable if the primordial curvature perturbations are enhanced at small scales.  
 In addition, the IGW spectrum is critically influenced by the equation of state of the universe at the time of their birth.  
 Actually, the detection of the relic GW stochastic background can serve as a probe of the very early  cosmic history, which is unknown for $t\lesssim 1$ s  \cite{Allahverdi:2020bys}. 
 In the following we will concentrate on GWs produced during radiation (RD) and early matter domination (EMD) eras.

 \subsection{ GWs in perturbation theory}

Let us consider a flat Friedmann-Robertson-Walker (FRW) background $g_{\mu\nu}^{(0)}=a^2(\tau) \eta_{\mu\nu}$. 
In the longitudinal or Newtonian gauge 
 the first-order scalar perturbations $(\Phi, \Psi)$ play the role of  gravitational potentials, and in the absence of anisotropic stress we have $\Phi=\Psi$. 
 In the second-order Einstein equations, there are combinations of the first-order  scalar perturbations $\Phi, \Psi$ which act as a source for the tensor modes identified as IGWs.
The evolution equation for the  IGW tensor $h_{ij}(\tau, \vec{x})$ is
\begin{equation} \label{inGW}
h''_{ij}+2{\cal H} h'_{ij}-\partial^2 h_{ij} \, = \, \Lambda_{ij, kl} F_{kl} \,.
\end{equation}
The rhs is the source term and $\Lambda_{ij, kl}$ is the tensor that projects the 
transverse-traceless part of the $F_{kl}(\tau, \vec{x})$.
The solution  in Fourier space is given by 
\begin{equation}
h_\lambda(\tau, \vec{k})=\frac{1}{a(\tau)} \int_0^\tau  G_k(\tau, \bar{\tau}, w) a(\bar{\tau}) F_\lambda(\bar{\tau}, \vec{k}) d\bar{\tau} \,,
\end{equation}
where $\lambda=+, \times$ and $e_{ij}^{(\lambda)}$ are the two polarization tensors.  $G_k(\tau, \bar{\tau}, w)$ is a Green's function which depends on the background equation of state, $w=p/\rho$. 
The Fourier mode of the source term is given by (see e.g. ref. \cite{Kohri:2018awv})
\begin{equation}
F_\textbf{k}=\int \frac{d^3 q}{(2\pi)^{3/2}} e_{ij}(\textbf{k})q_i \,q_j \left( 2\Phi_\textbf{q}\Phi_{\textbf{k}-\textbf{q}}+\frac{4}{3(1+w)}({\cal H}^{-1} \Phi_\textbf{q}+\Phi_\textbf{q})({\cal H}^{-1} \Phi_{\textbf{k}-\textbf{q}}+\Phi_{\textbf{k}-\textbf{q}}) \right) \,,
\end{equation}
where $F_\textbf{k}$ is the convolution of two first-order scalar perturbations at different wavenumbers and ${\cal H}=a H$.
The critical quantity is the  gravitational potential $\Phi$ that evolves according to the equation (for details see e.g. ref. \cite{Mukhanov:2005sc})
\begin{equation} \label{PhiEoM}
\Phi'' +3(1+c_s^2) {\cal H} \Phi'-c^2_s \nabla^2 \Phi+\left(2{\cal H}' +(1+3c^2_s){\cal H}^2\right) \Phi \propto 4\pi G a^2 \delta S \,.
\end{equation}
In the absence of entropy perturbations $ \delta S$  it has the exact solution
\begin{equation}
\Phi_k(\tau) =x^{-\ell} \left[ C_1(k)  J_\ell(x)   +C_2(k) Y_\ell(x) \right], \quad x\equiv \sqrt{w}k\tau, \quad \ell \equiv\frac{1}{2}\left(\frac{5+3w}{1+3w} \right) \,,
\end{equation}
where $J_\ell$ and $Y_\ell$ are Bessel functions of order $\ell$. For $w>0$ the potential decays once the perturbations enter the horizon and the IGW are generated mainly at the moment of horizon crossing. In particular,  for a RD era ($w=1/3$) the potential $\Phi(x)$ oscillates with a $x^{-2}$ decaying amplitude in subhorizon scales.  For an era of stiff-fluid domination, also called kination era,  $w=1$, the potential decays roughly as $x^{-3/2}$, see e.g. \cite{Domenech:2019quo, Dalianis:2020cla, Dalianis:2021dbs}.
However, for $w=0$ the potential does not decay and  $\Phi$ continuously feeds the induced tensors. 

The tensor power spectrum is expressed as a double integral involving the power spectrum of the curvature perturbations $\Delta^2_{\cal R}=\Delta^2_\Phi(5+3w)^2/(3+3w)^2$ \cite{Kohri:2018awv, Espinosa:2018eve}
\begin{equation}
{\Delta^2}_h(\tau, k)=4 \int_0^\infty dv \int_{|1-v|}^{1+v}du \left( \frac{4v^2-(1+v^2-u^2)^2}{4vu}\right)^2 I^2(v,u,x) {\Delta^2}_{\cal R}(kv) {\Delta^2}_{\cal R}(ku)
\end{equation}
where the dimesionless parameters are $x=k\tau$, $v=k_2/k_1$ and $u=|\textbf{k}_1-\textbf{k}_2|/ k$. The function $I(v,u,x)$ includes the time dependence.
The energy density parameter of IGW per unit logarithmic frequency interval  is given  in terms of the tensor power spectrum as $\Omega_{\textrm{IGW}}(\tau,k)\equiv{(1/24)}\ \left({k/\mathcal{H}(\tau)}\right)^2 {\Delta^2_{h}(\tau,k)}$.
As an  example, let us assume a Gaussian $\Delta^2_{\cal R}(k)$ with amplitude $A_{\cal R}$ and width $s$.  For  IGWs produced during RD, the spectral energy density parameter today at the frequency $f_\text{p}$ where $\Delta^2_{\cal R}$ is  maximal is  given by \cite{Dalianis:2020cla} 
\begin{align} \label{OmegaRDt0}
    \Omega^{(\text{RD})}_\text{IGW}(t_0, f_\text{p}) \sim \, 5 \times 10^{-9} \, s^2\, 
    \left(\frac{A_{\cal R}}{10^{-2}} \right)^2. 
\end{align}
A detailed analysis  can be found in \cite{Pi:2020otn}. Amplitudes ${\cal A}_{\cal R}\sim 10^{-2}$ can give detectable IGW signals, see e.g. \cite{Domenech:2021ztg} for a recent review.

 \subsubsection{Non-linearities}

 In an EMD era the pressure is negligible, $a\propto \tau^2$  and the linear evolution equation for $\Phi$ is given by 
\begin{equation}
\Phi''_k +\frac{6}{\tau}  \Phi'_k =0
\end{equation}
with the general solution $\Phi_k(\tau)=c_1(k)+{c_2(k)}/{\tau^5}$.  The gravitational potential and the density perturbation are related  through $\frac{\delta \rho_m}{\rho}=\frac{2}{3{\cal H}^2} \nabla^2_\text{com}\Phi$,
where $\nabla^2_\text{com}$ is  the comoving spatial Laplacian. For $\delta \rho/\rho \sim1$ and working in the Fourier space, one finds a scale $k^{-1}_\text{NL}$ at which  perturbation theory breaks down \cite{Assadullahi:2009nf}
\begin{equation}
k_\text{NL}(\tau) \sim 
\frac{{\cal H}(\tau)}{ {(\Delta^2_\Phi)}^{1/4} }\,.
\end{equation}
A perturbation $\delta \rho/\rho$  that has wavelength $k^{-1}$ and power $\Delta^2_\Phi$ becomes non-linear  at about  $(\Delta^2_\Phi)^{-1/4}$  Hubble times after horizon entry. 
Unless reheating is realized fast, one cannot track the evolution of the overdensities  during the EMD era in the framework of  perturbation theory. This fact restricts the perturbative analysis 
  only to wavenumbers $k<k_\text{NL}(\tau_\text{rh})$, where $\tau_\text{rh}$ is the conformal time of the cosmic reheating.   For example, for ${\Delta^2}_\Phi\sim10^{-9}$ one can only study scales with wavenumber  $k<{\cal O}(300) \, k_\text{rh}$, while  for ${\Delta^2}_\Phi\sim10^{-2}$ the scales shrink to wavenumbers  $k<{\cal O}(5) \,k_\text{rh}$. This means that  
the applicability of the perurbative formalism  is restricted only within a short period before the reheating of the universe.
 These  theoretical limitations call for a non-perturbative approach.

\subsection{ GWs beyond perturbation theory and during EMD }

A long period of EDM prior to BBN and a modification to the usually assumed RD  very early universe is common in many beyond the Standard Model scenarios. For example, moduli with masses around 100 TeV which decay through gravitationaly suppressed couplings lead to a late stage of reheating shortly before the time of BBN. 
During EMD,  IGW production can be described partly with   perturbation theory \cite{Inomata:2019ivs, Inomata:2019zqy}. Beyond the linear regime, 
a gravitational instability can be understood and traced in terms of the Zel'dovich solution \cite{Zeldovich:1969sb}.  It describes the  growth of inhomogeneities of nonrelativistic matter at scales withing the Hubble horizon 
and  without assuming a spherical symmetry.
In the absence of isotropic pressure it is natural to  assume   that the density perturbations have an initial small deviation from sphericity that we describe  by an ellipsoid.  In the Zel'dovich approximation the radial coordinate is 
$r_i = a(t)q_i + b(t)p_i(q_i) $, where $p_i$ is the deviation vector and $b(t)$ is a linearly growing mode in the EMD universe.  In an appropriate coordinate system we deal with the diagonal matrices $\partial p_i/\partial q_k=-{\rm diag}(\alpha, \beta, \gamma)$ and
$D_{i \ell} =\text{diag}(a-\alpha b, a-\beta b, a-\gamma b)$, where   $D_{i \ell}$ is  the deformation tensor.
The $(\alpha, \beta, \gamma)$ parameters describe the geometry of different overdensity configurations.

 GWs are emitted during the gravitational collapse because of the  aspherical and rapid motion of dense matter distributions.
The GW emission is described by a multipole expansion of the perturbation $h_{\mu\nu}$ on the  background spacetime $g^{(0)}_{\mu\nu}$. 
To lowest order  $h_{ij}(t, d) \propto \ddot{Q}_{ij}(t-d/c)/d$, where $Q_{ij}$ is the mass quadrupole moment and $d$ the distance from the source. 
This formula is valid for any nearly Newtonian slow-motion source. 
In terms of the moment of inertia the  quadrupole moment is written as
\begin{equation}
Q_{ij}=-I_{ij}(t)+ \frac13 \delta_{ij} \text{Tr} I(t),
\label{Qij}
\end{equation}
where $I_{ij}$ is the moment of inertia tensor defined as $I_{ij}=\int d^3r \, \rho ( \vec{r}) \left(\delta_{ij}|\vec{r}|^2- r_i r_j  \right)$.
After choosing  the principal axes frame, the moment of inertia tensor is $I_{ij}=\frac15 M {\rm diag}(r^2_2+r_3^2, r_1^2+r_3^2, r^2_1+r_2^2) $
where  $M$ is the constant mass enclosed by the ellipsoid.
The coordinates of the boundary evolve with time as
\begin{align}
r_i(t)  =\frac{3}{2}t_k^{1/3} t^{2/3}\left(1- \frac{\xi_i}{2}\left(\frac{t}{t_\text{max}}\right)^{2/3}\right),
\end{align}
where $\xi_i=\{1, \beta/\alpha, \gamma/
\alpha \}$ for the three directions respectively and $t_\text{max}=t_\text{max}(\alpha, \beta, \gamma, \sigma)$ is the moment of maximum expansion.  The variable   $\sigma$ stands for variance of the density perturbations.

\begin{figure}[t!]
\centering
 \includegraphics[width=0.48\textwidth]{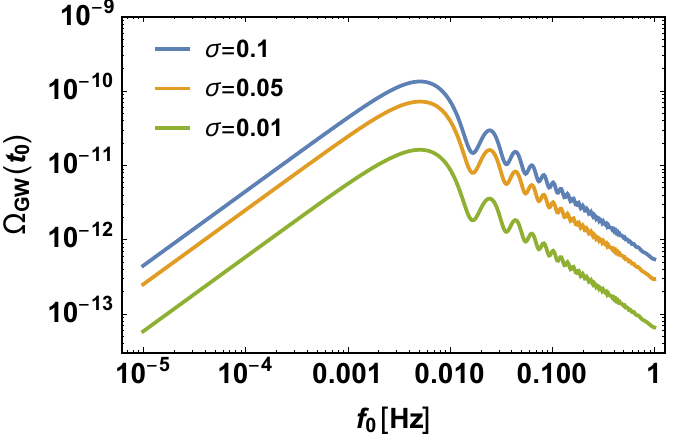}
\includegraphics[width=0.48\textwidth]{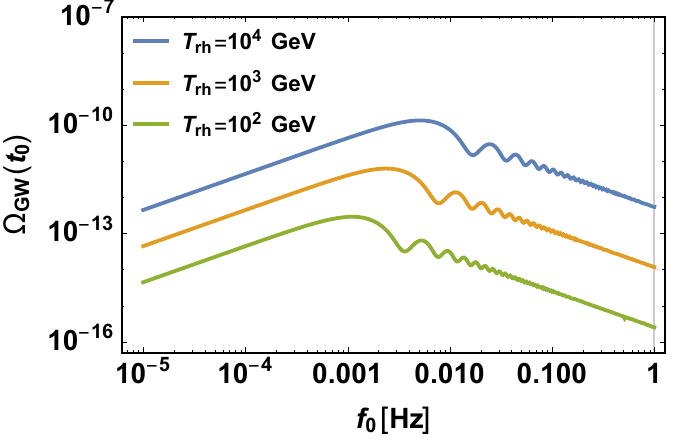}
\includegraphics[width=0.48\textwidth]{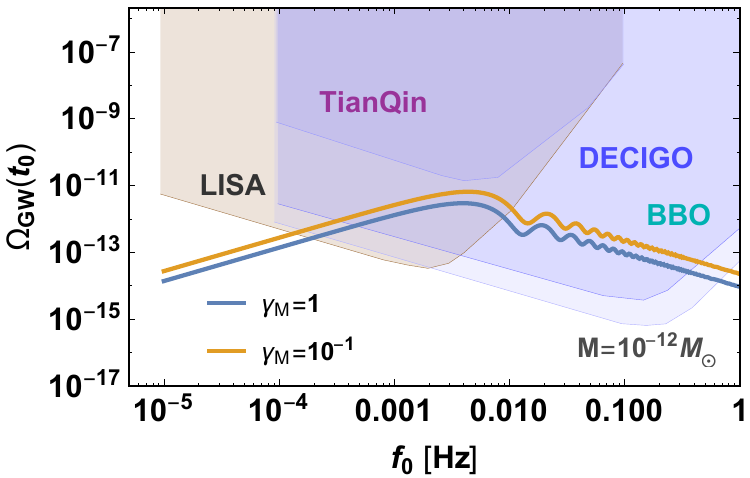}  \includegraphics[width=0.48\textwidth]{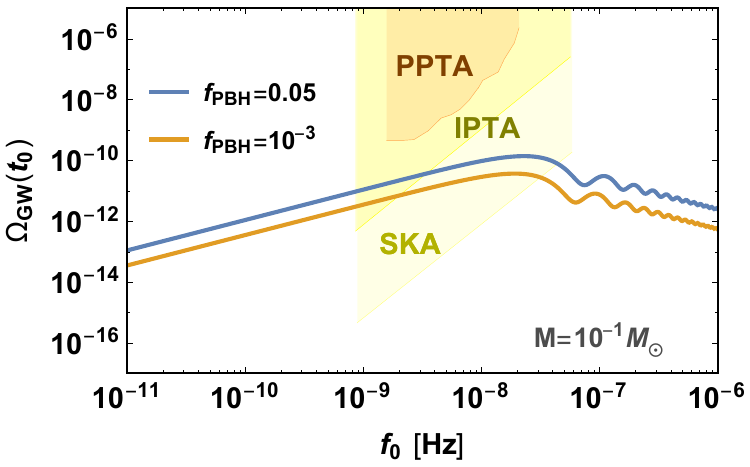}
\caption{
{\it Upper panels}: 
The  GW signal, eq. (\ref{result_Om}),  for a horizon mass $10^{-12} M_\odot$ for different $\sigma$ values and a reheating temperature $T_{\rm rh} = 10^4$ GeV (left panel) and for $\sigma=0.1$ and three different reheating temperatures (right panel).
{\it  Lower panels}:
The maximum GW signal in the LISA frequency band for the PBH dark matter scenario and  for two $\sigma$ values
 (left panel). The GW signal in the PTA frequency range, for a horizon mass $M = 10^{-1}M_\odot$ and for two fractional abundances for the associated PBH, 5\% and 0.1\% respectively, and reheating temperature  $\sim 80$ MeV  (right panel).}
\label{figEMD}
\end{figure}

The power emitted in the form of GWs from a Hubble patch  enclosing a perturbation with a quadrupole tensor $Q_{ij}$ is
\begin{equation}
\frac{dE_e}{dt}=\frac{G}{5c^5} \sum_{ij} \dddot{Q}_{ij}(t)\dddot{Q}_{ji}(t).
\end{equation}
with $G$ denoting Newton's  constant and $c$ the speed of light.
The differential energy emitted per logarithmic interval is
\begin{align} \nonumber
& \frac{dE_e}{d\ln \omega}=\frac{4\pi G}{5c^5} \omega^7 \sum_{ij}|\tilde{Q}_{ij}(\omega)|^2,
\end{align}
where $\tilde{Q}_{ij}$ are the Fourier modes in the continuum limit.  This expression gives the amount of GW spectral power produced in the EMD era due to a density perturbation with wavelength $q=k^{-1}$ that enters the horizon and evolves non-linearly  before the reheating of the universe.
Taking the integral over all GW sources, which are all the Hubble patches with size $q$ within our current horizon, and having included a weight function ${\cal F}_D$,   we find the total energy of the gravitational radiation emitted during the non-perturbative processes of  {\it collapse and halo formation}.
The ${\cal F}_D$ function is a probability distribution function (PDF) for $\alpha, \beta,$ and $ \gamma$ for 
  density perturbation with a Gaussian distribution,  called Doroshkevich PDF.
The differential energy density parameter of the stochastic GW background per observed logarithmic frequency
${\rho_\text{crit}}^{-1} {d\rho_\text{GW}}/{d \ln \omega}$ is
\cite{Dalianis:2020gup}
\begin{align} 
\Omega_\text{GW}(t,\omega_0) =\frac{1}{\rho_\text{crit}(t)}
 & \int\int\int_{\cal S}
\,d\alpha d\beta d\gamma\, \frac{1}{1+z} \frac{4 \pi G}{5 c^5}  \sum_{ij}|\tilde{Q}_{ij}\left(\omega\right)|^2 
\, \omega^7
 \left(\frac{4\pi}{3}q^3\right)^{-1} {{\cal F}_\text{D} (\alpha, \beta,\gamma, \sigma)}\,,   \label{result_Om}
\end{align} 
where  $\rho_\text{crit}$ is the critical energy density of the universe,  $\omega=\omega_0 (1+z)$, $z$  is the redshift and $f_0=2\pi \omega_0$ is the frequency parameter today. GW spectra described by eq. (\ref{result_Om}) are displayed in fig. \ref{figEMD}. The integration takes place over an appropriate region ${\cal S}$ of the parameter space $(\alpha, \beta, \gamma)$.

The central quantity  is the quadrupole $Q_{ij}(t)$ and the way it evolves determines the amplitude and the spectral characteristics of the GW signal. 
The quadrupole can be computed analytically within the Zel'dovich approximation until the moment of maximum expansion and  with adequate accuracy until  the moment of pancake collapse \cite{Dalianis:2020gup}.
The interesting stage of the quadrupole evolution beyond the pancake collapse can be probed by utilizing $N$-body  numerical tools\cite{DK}.
This method provides valuable quantitative insight on the non-linear regime and illuminates the underlying physical processes.
The specification of the spectral details of GWs produced from an EMD enables a reliable test of the 
very early cosmic history, which is unknown for $t\lesssim 1$ s,
through the detection of a relic GW stochastic background.

\section{Conclusions}

We briefly reviewed basic properties of three  inflation mechanisms that  transiently realize departures from the slow-roll paradigm. They are characterized by  the following features, respectively:  step-like transitions, approximate inflection points and sharp turns. 
Such features can enhance significantly the spectrum of curvature perturbations, possibly triggering PBH production and filling the universe with detectable cosmic GWs. 
We  described  the evolution of the perturbations
 and presented the produced spectra through analytic expressions  and complementary  numerical means. 
We highlighted the size of the enhancement and the pattern of  oscillations around the peak.   
 We also outlined two particular types of embedding of inflationary models with features into  fundamental high-energy frameworks:  $\alpha$-attractors and  SUGRA.
 Finally, we discussed and demonstrated the  observational consequences that the excited primordial scalar spectra have on  the IGW spectra. We  considered the standard RD era as well as the scenario of an  EMD era, where  calculations beyond  the linear regime are required in order to describe the full spectrum of GWs.
 It is actually the detection of the stochastic IGW background  
 that can be viewed as a direct window 
 to the primordial power spectrum of curvature perturbations  at small scales
 and can be used as a critical discriminator among different inflationary models.

\section*{Acknowledgments}
I would like to thank Vassilis Spanos and Nikolaos Tetradis for
collaboration in the work that is reviewed here and for
a careful reading of the manuscript. 
I would also like to thank the organization committee of the 11th Aegean Summer School that took place in Syros,
Greece for inviting me to present this review.
This work was supported by the Hellenic Foundation 
for Research and Innovation (H.F.R.I.) under the  “First Call for
H.F.R.I. Research Projects to support Faculty members and Researchers and the procurement of high-cost research equipment grant”  (Project Number: 824) and
 by the Cyprus Research and
Innovation Foundation grant EXCELLENCE/0421/0362.


\end{document}